\documentclass{aa}  

\usepackage{hyperref}

\usepackage{natbib}
\bibpunct{(}{)}{;}{a}{}{,}

\usepackage[varg]{txfonts}
\usepackage{epsf,palatino,url,graphicx,wrapfig,array,setspace,amsmath,amssymb,fancyhdr,multirow,lscape,appendix,rotating}
\usepackage{graphics,epsfig}

\usepackage{longtable}
\usepackage{amssymb}
\usepackage{xcolor}
\usepackage{color}
\usepackage{lipsum}
\usepackage{textcomp}
\usepackage{threeparttable}
\usepackage{enumerate}

\usepackage{placeins}
\usepackage{float}

\usepackage[normalem]{ulem}
\usepackage{soul}
\usepackage{blindtext}

\newcommand{\ergs}[1]{$\times 10^{#1}$ erg s$^{-1}$}

\newcommand{\ltsima}{$\buildrel < \over \sim$}
\newcommand{\lsim}{\lower.5ex\hbox{\ltsima}}
\newcommand{\gtsima}{$\buildrel > \over \sim$}
\newcommand{\gsim}{\lower.5ex\hbox{\gtsima}}

\newcommand{\xmm}{\hbox{\it XMM-Newton}\xspace}
\newcommand{\nus}{\hbox{\it NuSTAR}\xspace}

\newcommand{\ulx}{NGC\,300\,ULX1\xspace}

\newcommand{\eqb}{\begin{eqnarray}}
\newcommand{\eqe}{\end{eqnarray}}

\begin{document}

\title{Investigating ULX accretion flows and cyclotron resonance in NGC 300 ULX1}

\author{F. Koliopanos\inst{1,2}\thanks{\email{fkoliopanos@irap.omp.eu}}
 \and  G. Vasilopoulos\inst{3,4}
 \and  J. Buchner\inst{5}
  \and C. Maitra\inst{3}
 \and F. Haberl\inst{3}}

\titlerunning{\ulx Broadband spectroscopy}
\authorrunning{Koliopanos et al.}

\institute{CNRS, IRAP, 9 Av. colonel Roche, BP 44346, F-31028 Toulouse cedex 4, France
          \and Universit{\'e} de Toulouse; UPS-OMP; IRAP, Toulouse, France 
          \and Max-Planck-Institut für Extraterrestrische Physik, Giessenbachstraße, 85748 Garching, Germany
          \and Department of Astronomy, Yale University, PO Box 208101, New Haven CT 06520-8101, USA
          \and Pontificia Universidad Católica de Chile, Instituto de Astrofísica, Casilla 306, Santiago 22, Chile}

  \date{Received --, --; accepted --, --}

  \abstract
  {}
   {We investigate accretion models for the newly discovered pulsating ultraluminous X-ray source (ULX) NGC 300 ULX1}
   {We analyzed broadband \xmm and \nus observations of \ulx, performing phase-averaged and phase-resolved spectroscopy. Using the Bayesian framework we compared  two physically motivated models for the source spectrum: Non-thermal accretion column emission modeled by a power law with a high-energy exponential roll-off (AC model) vs multicolor thermal emission from an optically thick accretion envelope plus a hard power-law tail (MCAE model). The AC model is an often used phenomenological model for the emission of X-ray pulsars, while the MCAE model has been recently proposed for the emission of the optically thick accretion envelope expected to form in ultraluminous ($L_{X}>10^{39}$\,erg/s), highly magnetized accreting neutron stars.  We combine the findings of our Bayesian analysis with qualitative physical considerations to evaluate the suitability of each model.  }
   { The low-energy part ($<$2\,keV) of the source spectrum is dominated by  non-pulsating, multicolor thermal emission.  The (pulsating) high energy continuum is more ambiguous. If modelled with the AC model a residual structure is detected, that can be modeled using a broad Gaussian absorption line centered at ${\sim}12$\,keV. However, the same residuals can be successfully modeled using the MCAE model, without the need for the absorption-like feature.
   Model comparison, using the Bayesian approach strongly indicates that the MCAE model -- without the absorption line -- is the preferred model.  }
   {The spectro-temporal characteristics of \ulx are consistent with previously reported traits for X-ray pulsars and (pulsating) ULXs. All models considered strongly indicate the presence of an accretion disk truncated at a large distance from the central object, as has been recently suggested for a large fraction of both pulsating and non-pulsating ULXs.  The hard, pulsed emission is not described by a smooth spectral continuum. If modelled by a broad Gaussian absorption line, the fit residuals can be interpreted as a cyclotron scattering feature (CRSF) compatible with a ${\sim}10^{12}$\,G magnetic field. However, the MCAE model can successfully describe the spectral and temporal characteristics of the source emission, without the need for an additional absorption feature and yields physically meaningful parameter values. Therefore strong doubts are cast on the presence of a CRSF in \ulx. }


\maketitle


\section{Introduction}
\label{sec-intro}


Accretion powered binaries are some of the most luminous objects in the Universe. Due to their high luminosity radiation pressure can become so strong that it exceeds that of the in-falling matter -- essentially inhibiting steady accretion. The limit at which this takes place is called the Eddington luminosity ($L_{\rm Edd}$), and it scales linearly with the mass of the compact object. In the past decades there have been numerous sources observed at luminosities exceeding $L_{\rm Edd}$ for a stellar black hole (i.e.~${\sim}10^{39}$\,erg/s for a ${\sim}10\,M_{\odot}$  black hole). These ultra-luminous X-ray sources (ULXs), were initially believed to harbor intermediate mass black holes \citep{1999ApJ...519...89C,2000ApJ...535..632M,2001MNRAS.321L..29K,2003ApJ...585L..37M}.    However, it was soon understood that the majority (if not all) of the ULX population can be powered by stellar mass BHs accreting at super-Eddington rates \citep[e.g.][]{2003ApJ...596L.171G,2004NuPhS.132..369G,2004MNRAS.349.1193R,2007MNRAS.377.1187P,2009MNRAS.393L..41K,2017mbhe.confE..51K}.
The remarkable recent discoveries of three pulsating ULXs \citep{2014Natur.514..202B,2016ApJ...831L..14F,2017MNRAS.466L..48I,2017Sci...355..817I} have established that prolonged super-Eddington accretion onto stellar-mass objects can be sustained and at least a few ULXs can be powered by highly magnetized neutron stars (NS), as evidenced by the presence of pulsations.

This unexpected discovery is perhaps not that surprising as the strong magnetic fields may actually provide the most effective mechanism for ``bypassing'' the Eddington limit and producing the observed luminosities of ULXs. Furthermore, building on past and present theoretical considerations \citep{1976MNRAS.175..395B,2017MNRAS.tmp..143M,2017MNRAS.468L..59K}, it has been proposed that the majority of known ULXs -- not just the few pulsating ones -- may be powered by highly magnetized neutron stars rather than black holes \citep[e.g.][]{2017A&A...608A..47K,2018arXiv180304424W}.

\ulx is the fourth system classified as a pulsar ULX (PULX) \citep{2018MNRAS.476L..45C}.
This remarkable system has probably one of the fastest spin-up rates ever observed.
The spin period ($P_s$) of the NS was measured to be ${\sim}362$\,s in December 2016, while a recent measurement of its spin was ${\sim}20$\,s on January 2018. 
\cite{2018MNRAS.476L..45C} reported a very high spin-up rate ($\dot{P}_{s}=-5.56\times10^{-7}~s~s^{-1}$) of the neutron star within the duration of the simultaneous \xmm and \nus observations. 
In this paper we report on the detailed spectral properties of the system based on the \xmm and \nus observations performed in 2016.
We performed phase-resolved spectroscopy on the broadband spectra and explored the source behavior in the context of accreting highly magnetised NSs -- such as Be-XRBs, which is the likely nature of this source -- but also in the context of supercritically accreting high-B NSs as has been recently explored by \cite{2017A&A...608A..47K} and \cite{2018arXiv180304424W}. 

More specifically we test the source behavior against the spectral and temporal traits predicted for the emission of the accretion column that is expected to form in high-B NSs accreting at high rates. This involves the predicted spectral shape of the emission \citep[e.g.][]{1981ApJ...251..288N,1985ApJ...299..138M,1991ApJ...367..575B,2004ApJ...614..881H,2007ApJ...654..435B}, its hardness as a function of pulse-phase as has been noted in several X-ray pulsars (XRPs) \citep[e.g.][]{2011A&A...532A.126K,2015A&A...581A.121M,2017A&A...601A.126V,2018A&A...614A..23K} and also the presence of non-pulsating components such as the thermal-like excess below ${\sim}1$\,keV, which can be attributed to the presence of a truncated accretion disk \citep[e.g.][and references therein]{2004ApJ...614..881H}. Since the source is also a ULX, we also investigate the spectral characteristics for pulsating ULXs predicted by \cite{2017MNRAS.tmp..143M} and observationally investigated by \cite{2017A&A...608A..47K}.

\section{Observations and data reduction}
\label{sec-observations}

To study the broadband spectral behavior of \ulx we considered the 2016 joint \nus (21-12-2016, ObsID: 30202035002) and \xmm (17-12-2016, ObsIDs: 0791010101,0791010301) observation. 
We calculated the phase of each event based on the NS  ephemeris (reference date:~MJD${=}$57738.65661582, P${=}$31.718\,s, $\dot{\rm P}{=}55.3{\times}10^{-8}$\,s/s).
We extracted background subtracted light curves and phase-resolved and phase-averaged spectra from both telescopes. Both phase-resolved and phase-averaged spectra were analyzed simultaneously for the telescopes covering a spectral band between 0.3-30\,keV.  

For the analysis of the two \xmm observations, we only considered the EPIC-pn detector \citep{2001A&A...365L..18S} which has the largest effective area of the three EPIC detectors, in the full 0.3-10\,keV bandpass, and had registered more than ${\sim}$80000 photons for each of the observations considered, providing sufficient statistics for continuum spectral fitting of both phase-resolved and phased-averaged analysis. The data were handled using the \xmm data analysis software SAS version 16.1.0. and the calibration files released\footnote{https://www.cosmos.esa.int/web/xmm-newton/ccf-release-notes} until December 13th, 2017. Following standard extraction guidelines we filtered out any potential high background flares by extracting 10$<$E$<$12\,keV light curves with a 100\,s bin size and placing appropriate threshold count-rates for high energy-photons. Using these, we removed any time intervals affected by high particle background.
The pn detector was operated in imaging mode during both observations. For the spectral extraction we considered a circular extraction region with a radius 35\arcsec centered at the Chandra source coordinates \citep{2011ApJ...739L..51B} and in order to avoid possible contamination from a neighboring point source. However, in both observations the source  was adjacent to a chip gap and we note that when part of the PSF lies in a chip gap, effective exposure and encircled energy fraction may be affected.
We followed the standard filtering and extraction practices provided by the \xmm Science Analysis System (SAS). 
The source spectrum was extracted using SAS task \texttt{evselect}, and the standard filtering flags (\texttt{\#XMMEA\_EP \&\& PATTERN<=4} for pn).  For the background region
we considered a 40{\arcsec} circle located in the same region of the chip as the source center, avoiding the copper ring and OoT events. SAS tasks \texttt{rmfgen} and \texttt{arfgen} were respectively employed to produce the redistribution matrix and ancillary file. 

For the analysis of the \nus data we used the version 1.8.0 of the \nus data analysis system (\nus DAS), and instrumental calibration files from CalDB v20180312. The data were calibrated and cleaned using the standard settings on the \texttt{NUPIPELINE} script, reducing internal high-energy background, and screening for passages through the South Atlantic Anomaly (settings SAACALC$=$3, TENTACLE$=$NO and SAAMODE$=$OPTIMIZED). Using the \texttt{NUPRODUCTS} routine we extracted phase-averaged source and background spectra and instrumental responses were then produced for each of the two focal plane modules (FPMA/B). The spectral products were extracted from a circular region  of 50{\arcsec} radius, and background was estimated from same size regions of blank sky on the same detector as the source, as far from the source as we could to avoid contribution from the point-spread function (PSF) wings. We applied standard PSF, alignment, and vignetting corrections. 

For the ${\chi}^{2}$ analysis, all spectra were regrouped so as not to oversample the instrument energy resolution by more than a factor of 3 and to have at least 25 counts per bin. For the Bayesian analysis (see Sect.~\ref{sec:Bayesian}) the spectra were unbinned and Cash statistics \citep{1979ApJ...228..939C} was employed. The analysis was performed using the {\tt Xspec} spectral fitting package, version 12.9.0 \citep{1996ASPC..101...17A} and the Bayesian X-ray analysis package \citep[][]{2014A&A...564A.125B}. 
We restricted the \nus spectra to energies below 30.0 keV as above this range the signal-to-noise ratio (S/N)\footnote{We define S/N based on the total number of source ($\rm N_{\rm src}$) and background ($\rm N_{\rm bg}$) events that are collected from areas with same size as: ($\rm N_{\rm src}/\sqrt{\rm N_{\rm src}+N_{\rm bg}}$)} fell below a value of 5. We also ignored all \nus channels below 3\,keV.

\section{X-ray spectral analysis}
\label{sec:analysis}

\begin{figure}
  \resizebox{\hsize}{!}{
      \includegraphics[angle=0,clip=]{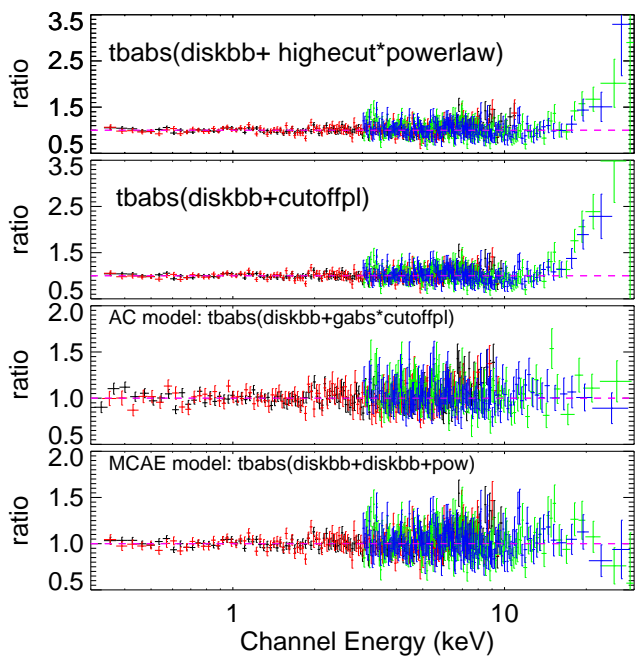}
 }
  \caption{Data-to-model ratio vs energy plot for the phase-averaged broadband (\xmm plus \nus) spectrum of \ulx. Top two panels: Absorbed power-law models with exponential roll-off (using two different \texttt{Xspec} models). Two lower panels: AC and MCAE models, respectively.} 
  \label{fig:ratio}
\end{figure}


\subsection{Accretion column emission}
\label{sec:spec}

PULXs are extremely luminous accreting NSs, thus it is natural to attempt to model their spectra in the context of Galactic accreting NS. 
In general the X-ray continuum emission of an XRP is composed by a pulsating hard component originating from the accretion column and a soft excess.
Thus, we consider an empirical model comprised of a soft multicolor disk black body (MCD) and a power law with a high energy roll-off, the MCD component is expected to originate in the truncated accretion disk and the power law simulates the accretion column emission \citep[e.g.][]{2007ApJ...654..435B}. We denote this model as the "Accretion column (AC) model".
We also note that as a first approximation a cutoff power-law (cPL\footnote{The photon distribution is ${\propto}E^{-\Gamma}\exp{-E/\beta}$, where $\Gamma$ is the photon index and $\beta$ the e-folding energy of the exponential roll-off)}) model can describe the X-ray spectra of non pulsating ULXs \citep{2017ApJ...836..113P}.

\subsection{The Multi-color Accretion envelope model for ULXs}
\label{sec_MCAE}

Alternatively, the characteristic roll-off in the pulsating and non-pulsating ULXs can be modeled with a hot thermal component that is believed to originate from an accretion envelope that engulfs the NS \citep{2017MNRAS.tmp..143M}. This envelope is produced by the free falling accreting matter that starts from the magnetospheric radius and extends to the top of the accretion column. For the extreme accretion rates required for PULXs the opacity of the envelope becomes high enough to reprocess most of the emission of the accretion column.
In this case,  the pulsed emission is the result  of the pronounced temperature gradient of the spinning accretion envelope whose spectrum is described by a hot (${\gtrsim}2$\,keV) multi-color thermal model. 

The broad band spectrum can then be modeled using a combination of two MCD components and a power-law tail \citep[][]{2017A&A...608A..47K}. The "soft" MCD is again used as a rough approximation of the truncated multi-temperature disk (however, the disk will most likely be geometrically thick, see \citealt{2017MNRAS.470.2799C} and also discussion in \citealt{2017A&A...608A..47K}). Similarly the second (hot) MCD component is an approximate description for the multicolor black body emission of the accretion envelope. The power-law tail is likely the result of up-scattering of photons by the free falling electrons outside the accretion column \citep[e.g.][]{1976SvA....20..436K,1988SvAL...14..390L,2013ApJ...777..115P} that form the accretion envelope. A considerable fraction of these non-thermal electrons are expected to escape the accretion column unprocessed, as discussed in \cite{2017MNRAS.tmp..143M} and \cite{2017A&A...608A..47K}.   We refer to this model as the "Multi-Color Accretion Envelope" (MCAE) model.

\begin{figure}
  \resizebox{\hsize}{!}{
     \includegraphics[angle=0,clip=]{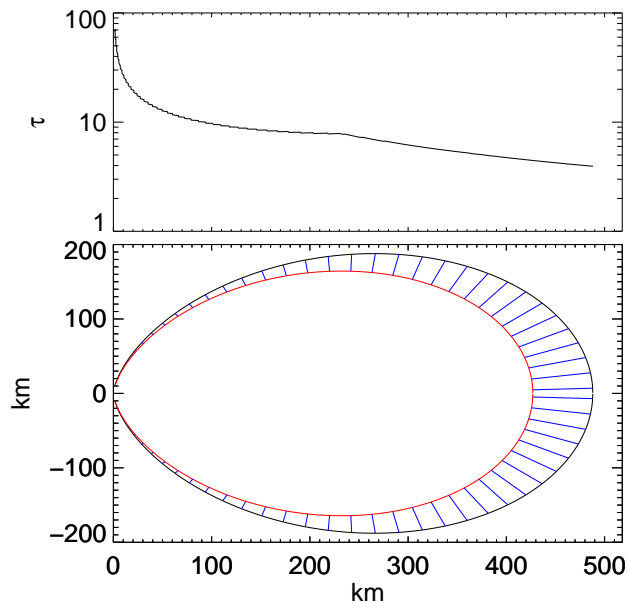}
 }
  \caption{Structure of the accretion envelope. Optical depth at a distance. across the envelope the optical depth has been computed based on the minimum separation of the two confining lines. The separation of the lines (i.e. the base of the accretion column) has been selected to be 100 m \citep[see discussion in][]{2017ApJ...835..129W}. 
  } 
  \label{fig:envelope}   
\end{figure}

Here, we briefly revisit the proposed model of accretion envelopes \citep{2017MNRAS.467.1202M,2017A&A...608A..47K} in order to apply its basic principles on \ulx and illustrate why this model very likely describes its accretion process.
We calculate the magnetospheric radius following \citet{2014EPJWC..6401001L} and assume that the total X-ray luminosity is equal to the rate at which gravitational energy of the in-falling matter is released ($L_x{\simeq}{GM\dot{M}/R_{NS}}$).
We assume that accreting matter follows the direction of the last closed magnetic field line (i.e. $R_{\rm m}$). 
Matter is then confined between a space defined by adjacent lines, this can be a small fraction of $R_{\rm m}$.  
At the base of the accreted flow a cylindrical accretion column is formed. Through diffusion processes the width of the accretion column can be broadened, but this does not affect our calculations especially away from the NS.
The equation of a dipole magnetic field line can be expressed in terms of its largest radius (e.g. $R_{\rm m}$) and its angular coordinate $(\theta)$ from the equatorial plane: $R(\theta)=R_{\rm m}\cos^2{\theta}$. For the last closed field line at the surface of the NS this corresponds to $\cos(\theta_{\rm NS})=\sqrt{R_{\rm NS}/R_{\rm m}}$ or $\theta_{\rm NS}{\sim}82^{\rm o}$. Thus we can express any magnetic line based on their trace on the NS surface by using the relation:
\begin{equation}
R(\theta)=R_{\rm NS}\frac{\cos^2{\theta}}{\cos^{2}{\theta_{\rm NS}}}
\label{eqn:line}    
\end{equation}
As a next step we need to calculate the size or width of the envelope at a given radius, and its minimum optical depth as a function of radius. 
By following two ''parallel'' closed field lines from the NS surface towards the magnetosphere we can calculate the surface area between them based on their minimum separation ($d_{\rm R,min}$) at a given point \citep[see ][for more details about the described configuration]{2017ApJ...835..129W}. Moreover, we calculate the surface area $S_{\rm D,R}$ of this conical frustum  that is defined by the orientation of $d_{\rm R,min}$ assuming axial symmetry around the magnetic axis (see Fig.~\ref{fig:envelope}). 
The opacity of the envelope in the direction of the minimum width can then be calculated as a function of radius (see also eq.~3 in \citealt{2017MNRAS.467.1202M}):
\begin{equation}
\tau(R)=\frac{\kappa_e \dot{M} d_{\rm R,min}}{2 S_{\rm D,R} \upsilon(R)},
\label{eqn:opt}  
\end{equation}
where $\kappa_e$ is the Thomson electron scattering opacity, and $\upsilon(R)$ is the local velocity of the in-falling matter, which cannot be larger than the free fall velocity. We note that at a given distance, from the rotational axis, the separation of two close field lines can only be a fraction of the distance thus the free fall velocity can vary by less than a factor of two within the range of the conical frustum. In Fig.~\ref{fig:envelope} we plot an example of the above defined envelope assuming different line configurations at the NS surface for the above paradigm, as well as the opacity derived by solving eq. \ref{eqn:opt}. The resulting magnetospheric envelope is closed and optically thick. 

We point out that in the case of oblique rotators, where the magnetic and rotation axis do not align a fully closed envelope could be difficult to maintain for large dynamical times. Most of the time material should fall into the polar cap from the direction where magnetic lines are closer to the inner disk radius, as indicated by simulations \citep{2017ApJ...851L..34P}. This would result in an envelope that covers only a fraction of the sphere around the NS. Given the above, the optical depth plotted in Fig.~\ref{fig:envelope} should only be considered as a lower limit.

\subsection{Modeling of the phase-averaged continuum}

Below we report on the two modeling approaches when applied to the pulse-phase averaged X-ray spectrum.

{\it AC model:} The \xmm EPIC-pn and \nus (FPMA and FPMB) phase-averaged spectra were fitted together with a constant multiplicative factor to account for inter-calibration uncertainties between the two instruments and any minor intensity
changes between the two XMM-Newton observations. The spectral continuum was modeled with an MCD with a temperature of ${\sim}0.29$\,keV (\texttt{ Xspec} model \texttt{ diskbb}) and a power law with a spectral slope of ${\sim}0.5$ and an e-folding energy of the exponential roll-off at $E_{fold}{\sim}4.2$\,keV (\texttt{ Xspec} model \texttt{ cutoffpl}, "Full (no line)" model in Table~\ref{tab:resolved}).
The interstellar absorption was modeled using the improved version of the \texttt{ tbabs} code\footnote{http://pulsar.sternwarte.uni-erlangen.de/wilms/research/tbabs/} \citep{2000ApJ...542..914W}. The atomic cross sections were adopted
from \cite{1992ApJ...400..699B}.  Our AC model is in principle a simplification of  the model used by \cite{2018MNRAS.476L..45C},  basically it is the \texttt{ powerlaw*highecut} model with $E_{cut}=0$, providing smoothest version of a cutoff power-law continuum with one less parameter than the \texttt{ powerlaw*highecut}. For completeness we have considered both power-law models in our full parameter space exploration (see below and also section \ref{sec:Bayesian}) but only tabulate the best fit values (Table~\ref{tab:resolved}) for the \texttt{ cutoffpl} model and the "AC model" only refers to this. For further simplicity, we  considered only a single instance of the \texttt{ tbabs} to account for the combined  Galactic foreground absorption and for both the interstellar medium of NGC 300, as well as the intrinsic absorption of the source (the partial absorption component is virtually insignificant during the 2016 observations, \citealt{2018MNRAS.476L..45C}).  Element abundances were taken from \cite{2000ApJ...542..914W} and cross-sections from \cite{1996ApJ...465..487V}. The model yielded a $\chi^2_{\rm red}$ value of 1.24 for 765  degrees of freedom (dof), with the data-to-model ratio  plot featuring some (minor) negative and strong positive residuals (Fig.~\ref{fig:ratio} second panel).  Modelling the continuum with the more general  \texttt{ powerlaw*highecut} model ($E_{\rm cut}=5.5\pm0.3$\,keV, $E_{\rm fold}=6.7\pm0.3$\,keV), yields a moderately better fit with a  $\chi^2_{\rm red}$ value of 1.19 for 764 dof. Similar (albeit slightly less pronounced) residuals appear in the \texttt{ powerlaw*highecut} data-to-ratio plot (see Fig.~\ref{fig:ratio}, top panel).  While the $\chi^2_{\rm red}$ values are not unacceptable the presence of systematic residual structure suggests that the smooth AC model,  on its own, cannot adequately describe the high energy ($>8$\,keV) part of the spectrum. 

{\it MCAE model:} We fit the spectral continuum with the MCAE combination of two MCD components and a power-law high energy tail.   Interstellar absorption was treated similarly to the AC model. The dual thermal model  yields a fit with reduced $\chi^2$ value of 1.09 for 764 dof, which is  an improvement of ${\Delta}{\chi}^2$ of 95 to the AC model. The bottom panel of Fig.~\ref{fig:ratio} presents the data-to-model ratio, where no prominent residuals are apparent.
 We note that the MCAE model {\it for the spectral continuum} has six free parameters and the AC model five free parameters.

\subsection{Phase-resolved spectroscopy}
\label{phase-res}

\begin{figure}
  \resizebox{\hsize}{!}{
     \includegraphics[angle=0,clip=]{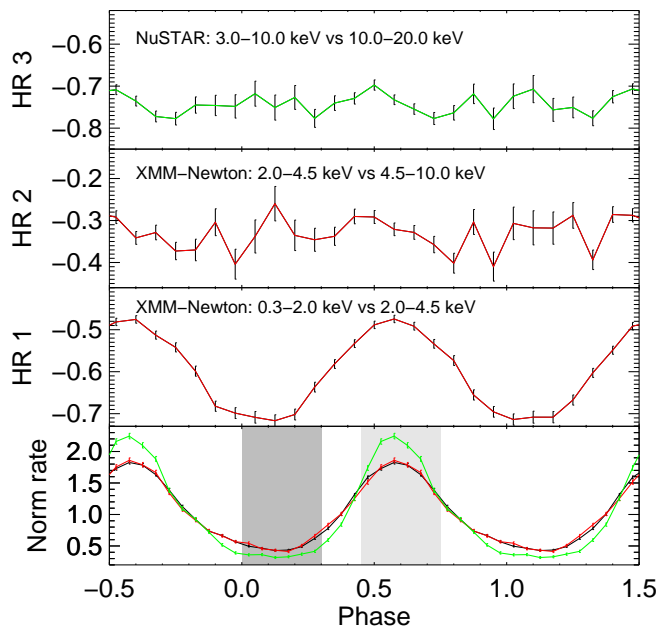}
 }
  \caption{Bottom Panel: Pulse profile of \ulx, using the 2016 \xmm/EPIC-pn (black, and red lines; 0.3-10.0\,keV) and \nus (green lines; 3.0-30.0\,keV). ON-Pulse and OFF-Pulse spectra were extracted from the gray shaded areas, which have a width of 0.3 in phase. Top Panels: \xmm  HR1: 0.3-2.0 vs 2.0-4.5 , HR2: 2.0-4.5 vs 4.5-10.0 , \nus HR3: 3.0-10.0 vs 10.0-20.0 } 
  \label{fig:pp}
\end{figure}

\begin{figure}
  \resizebox{\hsize}{!}{
     \includegraphics[angle=-90,clip=]{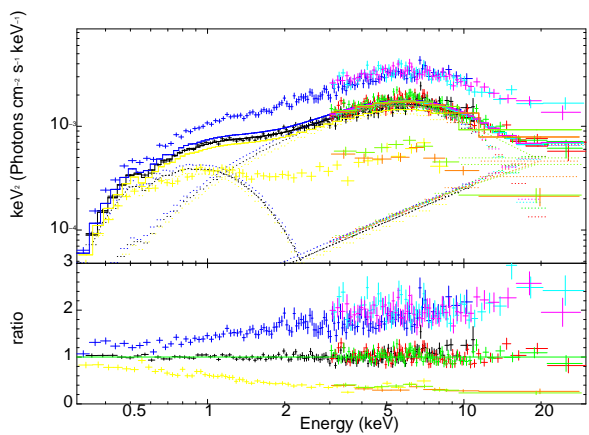}
 }
\caption{ Unfolded spectra and data-to-model ratio plot of the ON-Pulse, OFF-Pulse and FULL (phase-averaged) spectra from the MCAE modeling of the phase-averaged emission.}
 \label{fig:spec22}
\end{figure}

We followed \citet{2018MNRAS.476L..45C} to derive the pulse profile of the neutron star (Fig. \ref{fig:pp}).
To look for spectral changes with spin phase we studied the phase resolved hardness ratio (HR). We define HR as
$\rm{HR_{i}}=(\rm{R}_{i+1}-\rm{R}_{i})/(\rm{R}_{i+1} + \rm{R}_{i})$, with R$_{i}$ and R$_{i+1}$ denoting the background-subtracted count rates in two consecutive energy bands.
For the \xmm/EPIC-pn detector we used three energy bands (0.3-2.0 keV, 2.0-4.5 keV, 4.5-10.0 keV), while for \nus we used two energy bands (3.0-10.0 keV, 10.0-20.0 keV). Thus we defined three distinct HRs as shown in Fig. \ref{fig:pp}. From this first exercise it is clear that only the softer HR varies with phase. Specifically we see that HR$_{\rm 1}$ becomes softer when fainter, this behaviour is characteristic of the soft-excess seen in XRPs \citep[e.g.][]{2011A&A...532A.126K,2015MNRAS.452.1601P,2018A&A...614A..23K}.

Using the temporal information obtained from the source light curve we extracted phase-resolved spectra -- during the peak ("ON-Pulse") and the "trough" of the pulse profile ("OFF-Pulse"), see Fig.~\ref{fig:pp}. 
We studied the phase-resolved spectra by fitting them in a similar way as previously for the phase-averaged spectra.
The broadband ON-Pulse and OFF-Pulse spectra were fitted separately using the same models as in the phase-averaged spectrum. We monitored the behavior of the spectral components and the absorption, and the variation of their best fit parameters with phase. The resulting best fit values (i.e. median value of the posterior distribution and 90\% error bars) for the spectral modelling are presented in Table~\ref{tab:resolved}.  For purposes of presentation and to note the relative persistence of the soft emission vs the variability of the harder emission, we also unfolded the ON-Pulse and OFF-Pulse spectra simultaneously from a model frozen in the best fit parameters of the phase averaged (FULL) MCAE fit. In Fig.~\ref{fig:spec22} we present the unfolded spectra and data-to-model ratio vs energy plots to provide a qualitative assessment of this behavior.

\begin{table*}
 \caption {Central values and confidence levels  based on the marginalized parameter distribution of the Bayesian model comparison for the broadband fit of combined \xmm and \nus observation of NGC300 ULX-1, for phase-average (FULL) and phase-resolved spectra (ON- and OFF-Pulse). The continuum was fitted separately with the MCAE and the AC models. In both cases a Gaussian absorption line was used to model the possible {CRSF}. All errors are in the 90$\%$ confidence range.}
 \begin{center}
\scalebox{0.9}{   \begin{tabular}{lcccccccccccc}
     \hline\hline\noalign{\smallskip}
      MCAE model  & & \multicolumn{2}{c}{dBB soft}  & \multicolumn{2}{c}{dBB hot}& \multicolumn{2}{c}{PL tail}& \multicolumn{3}{c}{Gaussian line}\\
      \noalign{\smallskip}\hline\noalign{\smallskip}
      \noalign{\smallskip}\hline\noalign{\smallskip}
     \multicolumn{1}{l}{Phase} &
     \multicolumn{1}{c}{nH} &
     \multicolumn{1}{c}{k${\rm T_{disk}}$} &
     \multicolumn{1}{c}{${\rm {K_{disk}}}$} &
     \multicolumn{1}{c}{k${\rm T_{hot}}$} &
     \multicolumn{1}{c}{${\rm {{K_{hot}}}}$} &
     \multicolumn{1}{c}{${\rm \Gamma}$} &   
     \multicolumn{1}{c}{${\rm  {{K_{PL}}}}$} &  
          \multicolumn{1}{c}{${\rm E_{gabs}}$} &   
      \multicolumn{1}{c}{${\rm {\sigma}_{gabs}}$} &   
     \multicolumn{1}{c}{K$_{\rm gabs}$ }\\
     \noalign{\smallskip}\hline\noalign{\smallskip}
      
      \multicolumn{1}{c}{} &
      \multicolumn{1}{c}{[$10^{21}$\,cm$^{-2}$]} &     
      \multicolumn{1}{c}{keV} &
      \multicolumn{1}{c}{} &
      \multicolumn{1}{c}{keV} &
      \multicolumn{1}{c}{[${\times}10^{-3}$]} &
      \multicolumn{1}{c}{} &
      \multicolumn{1}{c}{[${\times}10^{-5}$]} &  
      \multicolumn{1}{c}{keV} &
      \multicolumn{1}{c}{keV} &
      \multicolumn{1}{c}{} \\
      \noalign{\smallskip}\hline\noalign{\smallskip}
       Full (no line)         & 0.82$_{-0.08}^{+0.09}$& 0.30$\pm0.013$          & 9.2$_{-1.8}^{+2.3}$ & 2.39$_{-0.04}^{+0.05}$ & 6.81$\pm$0.40 & 0.92$_{-0.50}^{+0.38}$ &1.8$_{-1.4}^{+3.2}$   & -- & -- & --   \\
             \noalign{\smallskip}\hline\noalign{\smallskip}
       Full         & 0.76$_{-0.08}^{+0.09}$& 0.32$\pm0.02$          & 7.09$_{-1.45}^{+1.96}$ & 2.74$_{-0.22}^{+0.92}$ & 4.37$_{-2.27}^{+1.29}$ & 1.09$_{-0.81}^{+3.02}$ &0.97$_{-0.96}^{+5.02}$   & 12.0$_{-1.0}^{+0.9}$& 3.2$_{-1.0}^{+1.6}$ & 2.9$_{-1.6}^{+7.0}$   \\
       ON-Pulse      & 0.88$_{-0.11}^{+0.15}$& 0.33$\pm0.03$          & 8.77$_{-2.82}^{+5.03}$ & 3.41$_{-0.89}^{+0.38}$ & 4.74$_{-1.24}^{+6.51}$ & 1.11$_{-0.98}^{+3.43}$ &0.27$_{-0.26}^{+5.02}$   & 13.6$_{-0.7}^{+0.5}$& 4.4$_{-1.6}^{+0.4}$ & 9.0$_{-6.8}^{+3.1}$   \\
       OFF-Pulse     & 0.62$_{-0.13}^{+0.14}$& 0.31$\pm0.02$          & 5.34$_{-1.44}^{+2.15}$ & 2.82$_{-0.35}^{+0.44}$ & 1.33$_{-0.48}^{+0.76}$ & 2.52$_{-2.20}^{+2.16}$ &0.002$_{-0.002}^{+0.39}$ & 11.2$_{-0.9}^{+1.7}$& 2.30$_{-0.24}^{+0.67}$ & 4.3$_{-1.5}^{+2.8}$   \\
       \noalign{\smallskip}\hline\noalign{\smallskip}
      \noalign{\smallskip}\hline\noalign{\smallskip}

       AC model & & \multicolumn{2}{c}{dBB soft}  & &  \multicolumn{3}{c}{cutoff PL}& \multicolumn{3}{c}{Gaussian line}\\
             \noalign{\smallskip}\hline\noalign{\smallskip}
      \noalign{\smallskip}\hline\noalign{\smallskip}
       \multicolumn{1}{l}{Phase} &
     \multicolumn{1}{c}{nH} &
     \multicolumn{1}{c}{k${\rm T_{disk}}$} &
     \multicolumn{1}{c}{${\rm {K_{disk}}}$} &
     \multicolumn{1}{c}{ } &  
     \multicolumn{1}{c}{${\rm  E_{fold}}$} &  
     \multicolumn{1}{c}{${\rm \Gamma}$} &  
     \multicolumn{1}{c}{${\rm {{K_{cPL}}}}$} &
          \multicolumn{1}{c}{${\rm E_{gabs}}$} &   
      \multicolumn{1}{c}{${\rm {\sigma}_{gabs}}$} &   
     \multicolumn{1}{c}{K$_{\rm gabs}$ }\\
     \noalign{\smallskip}\hline\noalign{\smallskip}
      
      \multicolumn{1}{c}{} &
      \multicolumn{1}{c}{[$10^{21}$\,cm$^{-2}$]} &     
      \multicolumn{1}{c}{keV} &
      \multicolumn{1}{c}{} &
      \multicolumn{1}{c}{} &
      \multicolumn{1}{c}{keV} &
      \multicolumn{1}{c}{} &
      \multicolumn{1}{c}{[${\times}10^{-5}$]} &  
      \multicolumn{1}{c}{keV} &
      \multicolumn{1}{c}{keV} &
      \multicolumn{1}{c}{} \\
      \noalign{\smallskip}\hline\noalign{\smallskip}
      Full (no line) & 0.86$_{-0.10}^{+0.09}$   & 0.29${\pm}0.02     $ & 11.2$_{-2.6}^{+3.0}$ & --         --   & 4.17$_{-0.22}^{+0.25}$ & 0.53$_{-0.08}^{+0.08}$ &    48.8$_{-2.8}^{+2.9}$    &      --             &     --              &  --                    \\
       \noalign{\smallskip}\hline\noalign{\smallskip}  
       Full         & 0.83$_{-0.06}^{+0.08}$& 0.30$_{-0.01}^{+0.02}$ & 9.63$_{-1.53}^{+2.35}$ & -- & 5.66$_{-0.32}^{+0.40}$  & 0.65$_{-0.06}^{+0.07}$ &46.3$_{-2.47}^{+2.79}$  & 13.1$\pm$0.4        & 3.7$\pm$0.3         & 5.1$\pm$1.0   \\
       ON-Pulse      & 1.10$_{-0.13}^{+0.14}$& 0.26$_{-0.01}^{+0.02}$ & 22.5$_{-6.91}^{+10.8}$ & -- & 6.05$_{-0.35}^{+0.64}$   & 0.78$_{-0.05}^{+0.07}$ &98.3$_{-3.16}^{+4.11}$  & 13.8$_{-0.3}^{+0.4}$& 3.26$\pm$0.24       & 5.3$_{-0.7}^{+0.8}$   \\
       OFF-Pulse     & 0.62$_{-0.12}^{+0.14}$& 0.31$\pm0.02$          & 5.60$_{-1.51}^{+2.47}$ &-- & 4.01$_{-0.72}^{+1.18}$   & 0.36$_{-0.26}^{+0.30}$ &13.2$_{-2.53}^{+3.04}$  & 11.6$_{-0.6}^{+1.2}$& 2.4$_{-0.3}^{+0.7}$ & 4.8$_{-1.6}^{+3.0}$   \\
             \noalign{\smallskip}\hline\noalign{\smallskip}
      \noalign{\smallskip}\hline\noalign{\smallskip}
      
    \end{tabular}   }
 \end{center}
  \tablefoot{{ ${\rm K_{disk}}$ is the normalization parameter for the \texttt{ diskbb} component. Namely ${\rm K_{disk}}$ = ${\rm(R_{\rm disk}/{D_{10kpc}})^{2}}\,\cos{i}$, where $R_{\rm disk}$ is the inner radius of the disk in km, ${\rm D_{10kpc}}$ is the distance in units of 10\,kpc and i is the inclination. ${\rm K_{hot}}$ is the same as ${\rm K_{disk}}$ but for the hot MCD component. ${\rm {{K_{cPL}}}}$ and ${\rm {{K_{PL}}}}$ are the power-law normalization parameters in units of photons/keV/cm$^2$/s at 1\,keV. {K$_{\rm gabs}$ } is the line depth. The optical depth at line center is K$_{\rm gabs}/{\rm {\sigma}_{gabs}}$/$\sqrt{2\pi}$. \\}  }
 \label{tab:resolved}
\end{table*}


\begin{table}
 \caption{Log-Evidence for the MCAE and AC models as described in Sect.~\ref{sec:Bayesian}.}
 \begin{center}
\scalebox{0.8}{   \begin{tabular}{lcc}
     \hline\hline\noalign{\smallskip}
          \multicolumn{1}{l}{AC model} &&\\
      \noalign{\smallskip}\hline\noalign{\smallskip}
      \noalign{\smallskip}\hline\noalign{\smallskip}
        Phase & $\log{\rm Z}$ & DOF \\
     \noalign{\smallskip}\hline\noalign{\smallskip}
     Full (no line) & -516.0$\pm$0.3 &  765\\
     Full (line)    & -467.0$\pm$0.3  &  762\\
     ON-Pulse (no line)     & -724.8$\pm$0.3  &  634\\
     ON-Pulse (line)     & -715.8$\pm$0.3  &  631\\
     OFF-Pulse    & -560.00$\pm$0.29  &  250\\
      \noalign{\smallskip}\hline\noalign{\smallskip}
            \noalign{\smallskip}\hline\noalign{\smallskip}
      \multicolumn{1}{l}{MCAE model} &&\\
      \noalign{\smallskip}\hline\noalign{\smallskip}
      \noalign{\smallskip}\hline\noalign{\smallskip}
        Phase & $\log{\rm Z}$ & DOF \\
     \noalign{\smallskip}\hline\noalign{\smallskip}
     Full (no line) & -460.62$\pm$0.28  & 764\\
     Full (line)    & -461.5$\pm$0.3  & 761 \\
     ON-Pulse (no line)    & -701.4$\pm$0.3  &  633\\
     ON-Pulse (line)    & -702.6$\pm$0.3  &  630\\
     OFF-Pulse    & -554.59$\pm$0.26  &  249\\
      \noalign{\smallskip}\hline\noalign{\smallskip}         

    \end{tabular}   }
     \end{center}
    
 \label{tab:evidence}
\end{table}

\subsection{Possible broad absorption line residuals}

\begin{figure}
  \resizebox{\hsize}{!}{
     \includegraphics[angle=-90,clip=]{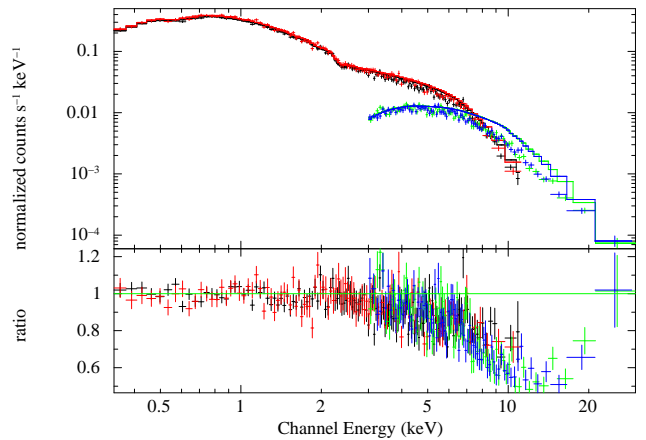}
 }
  \caption{Phase-averaged broadband (\xmm plus \nus) X-ray spectrum of \ulx (upper panel) plotted together with the data-to-model ratio for the absorbed AC model (lower panel). For illustration purposes the strength of the absorption line has been set to zero.} 
  \label{fig:cpl_noline}
\end{figure}

If we increase the $E_{\rm fold}$ value in the \texttt{cutoffpl} model -- in order to account for the high energy ($>$15\,keV) positive residuals -- we note the presence of broad negative residuals between ${\sim}8$ and ${\sim}$20\,keV (see Fig.~\ref{fig:cpl_noline}). Given the presence of a highly magnetised NS accretor, the broad absorption feature can be interpreted as a potential  cyclotron resonant scattering feature \citep[CRSF, e.g.,][]{2004AIPC..714..323H}. Modelling the broad absorption-like feature with a Gaussian absorption line (\texttt{Xspec} model \texttt{gabs}),  improves the quality of the AC model fit by a ${{\Delta}{\chi}}^{2}$ of 110 for three dof. As the third panel from the top of Fig.~\ref{fig:ratio} shows, the high-energy continuum is now adequately described.
The absorption line is centered at ${\sim}13.1\,$keV and has a width of ${\sim}3.7$\,keV (see Table~\ref{tab:resolved},  AC model, "Full").

Although a visual inspection of the residuals from the MCAE model does not reveal a need for an additional absorption feature (Fig.\ref{fig:ratio}, bottom panel), adding a Gaussian absorption line introduces a moderate improvement to the model (${\Delta}{\chi}^{2}$ of 12 for 3 dof) indicating the detection of a broad absorption line which is constrained at similar values ($E_{gabs}{\sim}$12\,keV and width of ${\sim}3$\,keV) to the AC model fit. However there is considerable degeneracy between line parameters and the parameters of the hot MCD and the power-law tail. There seems to be a region of parameter space for which the absorption line is not required by the MCAE fit.

What emerges  is a problem of comparison between two different -- but physically motivated -- models,  in an effort to determine whether the broad absorption-like feature can be incorporated in our modelling. A simple comparison between ${\chi}^2$ values cannot fully explore the multidimensional parameter space that needs to be investigated, in order to attain satisfactory constraints on the presence of the absorption feature. To pursue this task we employ a Bayesian model selection framework.

\section{Bayesian X-ray Analysis}
\label{sec:Bayesian}

To compare the two competing models, AC and MCAE, we use a Bayesian methodology \citep[see][]{2014A&A...564A.125B}. 
The evidence $Z$ is defined as
$$Z=\int\,\pi(\overrightarrow{\theta})\,\exp\left[-\chi^2(\overrightarrow{\theta})/2\right]\, d\overrightarrow{\theta},$$
where $\overrightarrow{\theta}$ is the parameter vector of the model, $\chi^2$ is the fit statistic and $\pi(\overrightarrow{\theta}) d\theta$ is the metric of the parameter space, which weighs the various regions (prior). In this work, we use uninformative priors throughout.

While commonly used model comparison methods include likelihood ratios at the best-fit or information criteria, here we prefer the model with the highest $Z$. Namely, the ratio  of $Z$ values from two models, the Bayes factor, can be used to assess their suitability. Intuitively speaking, $\log{Z}$ indicates the quality of the fit (i.e. $-\chi^2/2$) minus a penalty for size of the model parameter space. In Bayesian model selection, prediction diversity distant from the observed data is punished. 
The Bayesian approach is appealing because it does not assume the true value to be equal to the best fit and takes into account the uncertainties and the model complexity. While priors need to be specified on all parameters, primarily the priors on parameters that are different between models are influential.

To compute the evidence Z we used the Bayesian X-ray analysis package (BXA). The BXA package\footnote{BXA is available at https://github.com/JohannesBuchner/BXA} \citep[][]{2014A&A...564A.125B} connects the nested sampling \citep{2004AIPC..735..395S} algorithm MultiNest \citep{2009MNRAS.398.1601F} with {\texttt{Xspec}. It explores the parameter space and can be used for parameter estimation (probability distributions of each model parameter and their degeneracies) and model comparison (computation of $Z$).
We use BXA with its default parameters (400 live points and $\log Z$ accuracy of 0.1). 
By using this Bayesian framework we are able to test and compare the physically motivated spectral models -- described above -- with the X-ray spectrum of \ulx. Moreover we are able to test the variability of the spectral components with the spin pulse-phase.

\subsection{BXA results for \ulx}
\label{secbxares}

\begin{figure}
  \resizebox{\hsize}{!}{
     \includegraphics[angle=0,clip=]{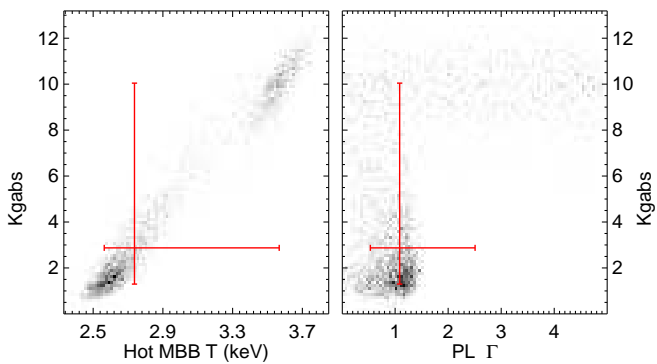}
 }
  \caption{Marginalised parameters of line depth ({K$_{\rm gabs}$ }, see Table \ref{tab:resolved} for details), temperature of hot thermal component and slope of power-law tail, for the MCAE model.   The red lines mark the 5\%, 50\% (cross-section) and 95\% percentiles of the 1D distributions. For the distributions of Fig.~\ref{fig:BXA_hist} we have plotted the 2D histograms for the two pairs of parameters (for the full parameter space see Fig.~\ref{fig:BXA_mcd}). } 
  \label{fig:bxaMCAE}
\end{figure}

We run the BXA routine for the entire set of models and for the phase-averaged and resolved spectra, presented in Sect.~\ref{sec:analysis}. We present the resulting $\log{Z}$ values in Table \ref{tab:evidence}.

Our analysis confirms that the AC model with a Gaussian absorption line is preferable to the one without, as both phase-averaged and resolved spectra have a significantly higher $Z$ value when modeled with the AC model that includes the feature. However, we further find that both the MCAE models with and without the line provide a better fit to the data, with the MCAE model without the line providing the superior fit to the phase-averaged ("Full") data-set, of all models considered. Namely, the "Full (no line)" MCAE fit (see Table \ref{tab:resolved}) yields a $\log Z$ value of ${\sim}-460.6$ and with a ${\Delta\log Z}$ of ${\sim}6.4$ from the AC model with the Gaussian absorption line and a ${\Delta\log Z}$ of ${\sim}1.0$ from the MCAE model with the line, it is the significantly preferable model. Based on the global evidence  the MCAE model with the Gaussian absorption line is about 25 times less likely than the simple MCAE fit.  The same results are  generated for the "ON-Pulse" and "OFF-Pulse" spectra as well (see Table \ref{tab:evidence}). In the following we focus on the analysis of the phase-averaged spectra as they have the best statistics, allowing for the significant detection of the absorption-like feature in the "AC model".

In Table \ref{tab:evidence} we tabulate the $\log{Z}$ values of the two continuum models with or without the absorption feature for the phase-averaged data-set (which has the best statistics to constrain the presence or absence of the line) and the  $\log{Z}$ values of the two continuum models with the absorption line for the ON and OFF-Pulse data-sets. Finally, we used the BXA to also compare between the \texttt{cutoffpl} and the \texttt{powerlaw*highecut} models. The algorithm confirms that the presence of the gaussian absorption line is also preferred in the \texttt{powerlaw*highecut} model. There is no significant difference in the preference between the two models, although it appears that without the absorption feature the \texttt{powerlaw*highecut} model yields a marginally better $\log{Z}$ value, but when including the Gaussian absorption feature the \texttt{cutoffpl} (AC model) is preferable. In all cases the MCAE model without the absorption line is the preferred model (see Table \ref{tab:evidence2} in Appendix A). We note that we are comparing models with a different number of free parameters (e.g.~the AC model {\it with} the Gaussian absorption has two additional free parameters with respect to the MCAE model that does not require it). Nevertheless, the Bayesian model comparison, punishes the increase in prediction diversity, caused by any additional free parameter(s). Therefore, the BXA is a more reliable method for model comparison and can more efficiently identify irrelevant parameters. For more detailed discussion of this point, the reader is referred to (e.g.) Section 2.4 of \cite{2018MNRAS.480.2377B}.

Closer exploration of the parameter space of the MCAE fit revealed a well defined region, for which the emission line becomes un-detectable.  The region is defined by lower temperatures for the hot thermal component (i.e.~k${\rm T_{hot}{\lesssim}2.4}$\,keV, see Fig.~\ref{fig:bxaMCAE}, left panel) ) combined with low values of the power-law spectral index (see Fig.~\ref{fig:bxaMCAE}, right panel), in agreement with the values expected for ULXs within the MCAE paradigm \citep{2017MNRAS.tmp..143M,2017A&A...608A..47K}. 
We also note that the tabulated values in Table \ref{tab:resolved} are not the best fit values of a single spectral fitting, but the posterior distribution of each model parameter in the 90\% confidence range. The central values are the median of the distribution. It is important to stress that -- while in a Gaussian distribution the median would coincide with the most likely value of the parameter -- this is not the case here, especially in the MCAE model.

In our scheme the hot MCD model is used as a rough approximation of the multi-temperature thermal emission of a spheroidal accretion envelope, therefore the size of the value of the inner radius that corresponds to the tabulated value of $K_{\rm disk}$ is physically meaningless \citep[see discussion in][]{2017A&A...608A..47K}. Alternatively, we can also model the envelope emission with a spherical black body model (\texttt{Xspec} model \texttt{bbodyrad}).  This model yields a radius of $49.5_{-10.8}^{+11.0}$\,km for a temperature of $1.56{\pm}0.03$\,keV (see Table~\ref{app2noline}). The \texttt{bbodyrad} fit returns a reduced $\chi^2$ of 1.11 (a ${\Delta}{\chi}^2$ increase of 8 for the same dof as the \texttt{diskbb} fit).  The increase may be an indication that the hot thermal emission is, indeed, better described by a multi-temperature thermal component, instead of a simple black body. However the $\chi^2$ of the \texttt{bbodyrad} fit is within an acceptable range and further comparison between the two components is beyond our scope. The  \texttt{bbodyrad} fit is only used to provide a rough estimate of the size of the emitting region, using a more ``physical'' parameter (i.e.~size of a thermally emitting sphere). The best fit values and the MCAE fit using both the \texttt{diskbb} and \texttt{bbodyrad} models are tabulated in Table~\ref{app2noline} in the appendix (we remind the reader that the main Table (1) presents the central values and confident levels of the model parameters from the BXA analysis).

\section{Discussion}

Modelling of the broadband emission of \ulx reveals a spectrum that shares similarities with both the spectra of accreting XRPs and ULXs. The spectral continuum is comprised of two distinct components: A soft thermal-like excess and a hard power-law like component with a high-energy rollover. We find that the presence of considerable residual structure in the high energy ($>8$\,keV) spectral continuum of the source can be modeled using a broad Gaussian absorption feature which can be interpreted as a CRSF. If confirmed by further  observations this is the first cyclotron scattering feature detected in an accretion powered pulsar at this luminosity regime and can have profound implications on our understanding of super-Eddington accretion onto high-B NSs and ULXs in general. However, we also find that when the spectral continuum is modeled using recent models for the spectra of pulsating ULXs \citep{2017MNRAS.tmp..143M,2017A&A...608A..47K} the absorption feature is not required. Below, we discuss the two models we have used to fit the spectral continuum of \ulx, we present the different emission components we have identified in the source spectrum and investigate their significance and behavior with pulse phase.

\vspace{0.5cm}

\subsection{Soft thermal component}

In all our spectral fits  we significantly detect the presence of a soft thermal component, which can be described well by a  ${\sim}0.3$\,keV multi-color disk with an inner radius between 400 and 800\,km (see Table \ref{tab:resolved} for the MCAE and AC models and assuming an inclination angle of 60${\degr}$). The characteristics of the soft thermal component are consistent with a geometrically thin accretion disk truncated at the magnetosphere of a dipole magnetic field with an equatorial strength of ${\sim}$3-7${\times}10^{12}$\,G, assuming the expressions in \citeauthor{2014EPJWC..6401001L} (2014; see also eq.~1 from \citealt{2017MNRAS.tmp..143M}). We note that this is only a rough estimation that suggests a magnetic field strength that is higher than $10^{12}$\,G. Different disk inclinations (e.g.~a few degree inclination as is suggested from the upper limit in the orbital plane in \citealt{2018MNRAS.476L..45C}) can  affect the B-value by a factor of ${\sim}$2, if we further take into account the color correction expected for the spectral emission of a disk of $>$200\,eV temperature (see \citealt{1986ApJ...306..170L,1986SvAL...12..383L,1995ApJ...445..780S} and Sect.~3 in \citealt{2017A&A...608A..47K}), we expect a factor more than six increase in the B-value, thus exceeding $10^{13}$\,G.

The column density and the emission intensity vary over the pulse phase. They appear to be positively correlated and peak during the ON-Pulse phase. This can be seen in the $L_{\rm x}$ and $N_{\rm H}$ panels in Fig.~\ref{fig:BXA_hist} (first panel) and Fig.~\ref{fig:BXA_hist_MCD}, for the AC and MCAE fits, respectively. The correlation is more pronounced in the AC model. However, we must note that there is considerable degeneracy between the soft thermal emission, the value of the $N_\mathrm{H}$ and the tail of the pulsating, hard component. Namely, in the AC model the non-thermal emission of the accretion column should also feature a low-energy roll-over, characteristic of the thermal or Bremsstrahlung seed photons that generate it. However, such a component is not included in any of the simple power-law models we considered.
As the flux of the non-thermal component varies with pulse phase (see discussion below), this will also affect the low energy part of the spectrum.  The variability of both the soft thermal component and the $N_\mathrm{H}$ is most likely due to this effect, rather than an inherent source characteristic (see also discussion in \citealt{2018A&A...614A..23K}). 
This problem is less pronounced in the MCAE model, since the \texttt{diskbb} model, used to fit the hot thermal emission of the accretion envelope, exhibits the Rayleigh-Jeans tail at low energies.

\subsection{Accretion column model}

The high energy part of the source continuum was modeled using a hard power-law with a spectral index of ${\sim}0.7$ and $E_{\rm fold}$ of ${\sim}6\,$keV (with the addition of Gaussian absorption line, see Table~\ref{tab:resolved} -- AC model "Full" and Sect.~\ref{cyclo} below). This spectrum is qualitatively similar to the spectra expected for the emission of the accretion column in highly magnetized NSs \citep[e.g.][]{1983ApJ...270..711W,2007ApJ...654..435B}. However, the energy of the cutoff is detected at an unusually low energy ($E_{\rm cut}{\approx}6$\,keV, see Sect.~3.1 and also \citealt{2018MNRAS.476L..45C}), compared to $E_{\rm cut}{\sim}$20-30\,keV  observed in most nominal X-ray pulsars \citep[e.g.][]{1983ApJ...270..711W,1998ApJ...509..897D,2002ApJ...580..394C,2013ApJ...775...65M,2013ApJ...779...69F,2014ApJ...780..133F,2017ApJ...841...35F} and also predicted by thermal and bulk comptonization models in the accretion column \citep[e.g.][]{2007ApJ...654..435B}. On the other hand the observed roll-off energy of \ulx is similar to that of most ULXs \citep[e.g.][]{2006MNRAS.368..397S,2007Ap..SS.311..203R,2009MNRAS.397.1836G,2013MNRAS.435.1758S}. 

The absolute flux of the cPL component clearly changes with pulse phase, with a value of $10.5{\pm}0.2{\times}10^{-12}$\,erg\,cm$^{-2}$\,s$^{-1}$ during ON-Pulse and $1.7{\pm}0.1{\times}10^{-12}$\,erg\,cm$^{-2}$\,s$^{-1}$ during OFF-Pulse. This behavior can also be evidenced by the flux ratio between the soft thermal component and the cutoff power-law which is, $F_{\rm cPL}/F_{\rm disk}=7.3\pm0.5$ during ON-Pulse and $F_{\rm cPL}/F_{\rm disk}=2.2\pm0.8$ during OFF-Pulse  (see also Fig.~\ref{fig:spec22}). This correlation indicates  that the non-thermal emission is the source of the pulsations. It can, therefore, be argued that the power-law component originates in the accretion column of a high-B NS. 

The non-thermal emission in X-ray pulsars often exhibits an anti-correlation with the pulse amplitude, such that the continuum/power-law becomes harder near the pulse peak \citep[e.g.][]{2011A&A...532A.126K,2015A&A...581A.121M,2017A&A...601A.126V,2018A&A...614A..23K}. 
In the context of an accretion column the hardening of the slope can be understood in terms of change in the inclination angle at which the observer views the column during ON and OFF-Pulse. Assuming that the accretion column is viewed closer to the face-on orientation during the ON-Pulse phase, it is also reasonable to assume that the scattering optical depth will increase, resulting in a harder slope. This can be understood in terms of the Comptonization y-parameter and the scattering of low-energy photons at high optical depths -- expected in the accretion column. Namely, comptonized photons with $h{\nu}{\ll}kT_{\rm e}$, are expected to follow a power-law distribution with $F(E)=C\,E^{-\Gamma}$, where $E$ is the photon energy \cite[e.g.][]{1983ASPRv...2..189P}, where ${\Gamma}=-3/2+\sqrt{9/4 + 4/y}$ \citep[e.g.][]{1979rpa..book.....R}, and $y$ is the Comptonization y-parameter, which in the presence of a strong magnetic field and high optical depth will be given by $y=2/15\,(kT/m_{\rm e}{c}^{2}){\tau}^2$ \citep{1975A&A....42..311B}. Therefore, as the optical depth increases the power-law emission is expected to become harder.

This behavior is not observed in \ulx. The median fitted power-law slope in the ON-phase is higher than the OFF-Phase, within their fairly large 90\% error-bars. From (see Table ~\ref{tab:resolved} and Fig.~\ref{fig:BXA_hist}), both values are consistent with each other, but also allow a change in $\Gamma$ by ${\sim}$0.3 between ON and OFF phase.  The uncertainty in the value of the spectral index is partially due to degeneracy between the index, the cutoff energy and the cross-calibration parameters between the detectors of the two telescopes. Indeed, when freezing the cross-calibration parameters, the BXA yields a value of  ${\Gamma}=0.63\pm0.10$ for the ON-Pulse spectrum and  ${\Gamma}=0.68\pm0.15$ for the OFF-Pulse spectrum. The absence of pulse-correlated hardening is further evidenced in the 3-10\,keV / 10-20\,keV hardness ratio vs pulse phase in Fig.~\ref{fig:pp}.

\subsection{A possible cyclotron line}
\label{cyclo}

The smooth AC model cannot adequately fit the ${>}$10\,keV range of the emission  unless a broadened absorption line is added (see Fig. \ref{fig:ratio}). In the context of a highly magnetized NS, this feature can be interpreted as electron cyclotron scattering.
The broad and shallow absorption feature in \ulx differs from the recently suggested narrow and deep feature in M51 ULX-8, which favors a proton cyclotron line \citep{2018arXiv180512140B}.
The identification of the line centroid energy is a robust method to estimate the magnetic field strength of the scattering region in a NS. The approximate relation between the observed electron cyclotron energy and the magnetic field strength in the scattering region is given by:
\begin{equation}
	E_{\mathrm{cyc}} = \frac{11.57\,\mathrm{keV}}{1+z} \times B_{12}
	\label{eqn:12b12}
\end{equation}
where $B_{12}$ is the magnetic field in units of $10^{12}$\,G and $z \sim 0.15$ is the gravitational redshift for standard NS parameters. If the presence of the cyclotron line is true, a line centroid at $\sim$ 13 keV implies a magnetic field strength of $B\sim10^{12}$ G. However, this magnetic field estimate may not reflect the surface magnetic field strength of the NS as the accretion column (and hence the line forming region) of an ULX may be formed several kms above the surface depending on the mass accretion rate and the surface magnetic field strength \citep{2015MNRAS.454.2539M}. Furthermore, magnetar-like field strengths are expected for ULXs as the average electron scattering cross-section is significantly reduced at high enough magnetic field strengths implying higher peak luminosities. Therefore the maximum accretion luminosity can be used to constrain the surface field strength of the NS as $B_{12}{\gtrsim}4\,L^{4/3}_{39}$  \citep{2014EPJWC..6402005M}. In the case of \ulx, a peak luminosity of 4.5\ergs{39} (0.3-30.0 keV) implies a surface magnetic field strength of ${\gtrsim}2\times10^{13}$ G. 

While this value is inconsistent with the centroid energy of the line, it is in agreement with estimations from accretion torque theory, based on the observed source spin-up rate (Vasilopoulos et al. submitted).
Assuming a dipolar magnetic field, the difference between the estimated field strength and the strength, derived from absorption features, imply a very tall accretion column of height $\sim1.7R_{NS}$.
However the dependence of the line on the continuum modelling makes it highly uncertain and needs to be confirmed by future deeper observations.  

Recent theoretical considerations argue that the cyclotron line may form when the radiation emitted by the accretion column is reflected from the neutron star surface \citep{2013ApJ...777..115P}. As accretion rate increases the accretion column grows larger and the illuminated fraction of the stellar surface becomes larger, weakening the average magnetic field, and reducing the cyclotron line energy.
If the accretion rate continues to increase (as the source luminosity persistently exceeds the Eddington limit) the height of the column will also increase, gradually reducing the fraction of its emission reflected of the NS-surface. As a result in the cyclotron reflection paradigm, very luminous sources will be expected to feature very weak cyclotron lines. The saturation of the reflected component is observationally indicated at $L_{X}{\sim}10^{38}$\,erg\,${\rm s^{-1}}$ \citep[e.g.][]{2015MNRAS.452.1601P}, above which the height of the accretion column is expected to exceed multiple times the radius of the NS and thus the reflected fraction is expected to become insignificant \citep[e.g. Fig.~2][]{2013ApJ...777..115P}.} 

The presence of a possible cyclotron line in \ulx was first reported  by \cite{2018arXiv180307571W}. In their work \citeauthor{2018arXiv180307571W} present the detection of the absorption-like feature, assuming the underlying continuum comprised of a cutoff power law and a soft thermal component. The authors  interpret the feature as a CRSF and then focus their discussion on the important consequences of this discovery. Using more thorough and refined statistical analysis, we confirm the findings of \citeauthor{2018arXiv180307571W} -- under assumptions similar to theirs -- but we also explore the possibility that the presence of this broad  absorption-like feature is contingent upon the choice of the underlying model. This caveat is acutely important, since the alternative model (i.e.~the MCAE model discussed below) is a physically motivated model, its best fit parameter values are consistent with the values found for other ULXs \citep[e.g.][]{2017A&A...608A..47K,2018arXiv180304424W} and does not require the addition of an absorption line.

\subsection{Multicolor accretion envelope model}

As the source luminosity lies well within the range of ULXs, we also modeled the spectrum using the MCAE model, based on the predictions of \cite{2017MNRAS.tmp..143M} and following the modelling scheme presented in \cite{2017A&A...608A..47K}. The continuum is described by two MCDs and a faint power-law tail. This model is similar to the one used for ULXs  in \cite{2017A&A...608A..47K} and the best fit values lie within the range found in that work. In the context of the MCAE model, the soft MCD component originates in the truncated disk -- similar to the AC model -- while the hot MCD component is interpreted as emission from an optically thick accretion envelope that is predicted to engulf the entire NS and almost entirely reprocess the emission of the accretion column \citep[see][and Sect.~\ref{sec_MCAE}]{2017MNRAS.tmp..143M}. 

The hot thermal component naturally explains the unusually low energy of the spectral roll-off, and can potentially reproduce the smooth, single-peaked pulsations which -- in the context of the MCAE -- are the result of an extended, rotating hot, optically thick emitting region that heavily distorts the original pulse-profile of the accretion column \citep[][]{2017MNRAS.tmp..143M}. Finally the presence of a fainter, hard power-law tail is similar to the ones detected in all ULXs with available \nus spectra in \cite{2017A&A...608A..47K}. More importantly the MCAE fit does not require the presence of the Gaussian absorption line.
 
The analysis with BXA revealed two different parameter solutions for the MCAE model, that can be easily evidenced in Figures \ref{fig:bxaMCAE} and \ref{fig:BXA_mcd}. In one solution (presented in Table~\ref{tab:resolved}) the broad absorption-like feature can be detected -- moderately increasing the logZ value -- while in the second  it is not detected (see Fig.~\ref{fig:BXA_mcd}). The model comparison using the logZ value favors the MCAE model (with or without the Gaussian absorption line) over the AC model (with the inclusion of the line). Furthermore, the preferred solution out of all models is the MCAE model without the Gaussian absorption line. This finding casts legitimate doubts on the robustness of a CRSF discovery. 

We note that in this model -- which is based on fundamental principles -- at the mass accretion rate implied by the observed luminosity, the accretion shell becomes partially or fully optically thick (see eq. \ref{eqn:opt} and Fig.~\ref{fig:envelope}). In addition to the shape of the pulse profile, the presence of an optically thick accretion envelope  would also explain the fact that the pulsed component does not become harder during the ON-Pulse phase. Pulsations -- in this scenario -- naturally arise from the rotation of the accretion funnel, as the observing angle of its surface would change with pulse phase.
This is in contrast to the explanation given by \cite{2018arXiv180307571W}, where the authors attributed the diluted pulsed profiles of PULXs, in general, to the presence of quadrupolar magnetic field geometry. Given the dimensions of the accretion column and the optical thick funnel/envelope around it contradicts their assumption that emission originates from near the NS surface.  Both features are more consistent with emission from a hot region on a rotating area, rather than anisotropic non-thermal emission from an accretion column. Therefore, the presence of a dual thermal spectrum is a likely scenario and the use of the MCAE model is physically motivated.

Since the inner disk radius provided by the MCD model -- which we used to fit the hot thermal emission -- lacks a physical meaning in the context of a spheroidal envelope, we consider the value of the spherical black body (BB) fit presented in Sect.~3. The BB model yields a size of $R_{\rm BB}=50{\pm}11$\,km.  As a simple spherical black body is a simplification of the predicted multicolor emission of the accretion envelope, this estimate can be viewed only as indicative of the size of the emitting region, which lies in the ${\sim}$100\,km range. The estimated size is smaller than the size of the magnetosphere, inferred from the soft MCD component (see Fig.~\ref{fig:BXA_hist_MCD} panel 2, top row) and, since most of the hot thermal emission originates from the hottest region of the accretion envelope, we can assume that $R_{\rm BB}$ is the size of this extended emission region, which occupies only a fraction of the optically thick accretion envelope.  Lastly, considering the uncertainty in the source inclination we can reasonably argue that the value of $R_{\rm BB}$ and the inner disk radius are in rough agreement. If we further assume that the CRSF can be detected in the accretion envelope paradigm, and assuming that it is formed at a distance of 50\,km from the NS surface, we infer a magnetic field strength of $5-6\times10^{13}$\,G on the surface of the NS.

This is in rough agreement with the  luminosity-based estimate and the estimate from accretion torque theory (Vasilopoulos et al. submitted). However, we must stress that -- (particularly) within the paradigm that wants the cyclotron line to be formed as a result of NS surface reflection \citep{2013ApJ...777..115P} and for the MCAE model -- the (weak) line would most likely not be detected due to obscuration of the central source. This would explain why within this framework a cyclotron line is indeed not significantly detected.

\subsection{Physical interpretation of fit derived quantities }

Measurement of the size of the emitting region can also provide a consistency/validity check between the AC and MCAE models. The fundamental difference of the two models is that the AC model predicts non-thermal (cPL) emission from the accretion column as the source of the pulsations, while the MCAE predicts hot thermal emission from an optically thick envelope. 
In both cases a characteristic effective temperature $T_{\rm eff}$ can be derived by the observed $L_X$, assuming a size for the emitting region (i.e. $L_{X}\propto R^2 {\sigma}T_{\rm eff}^4$).
Moreover, $3kT_{\rm eff}\lesssim E_{\rm \rm cut}$ and thus,  the size of the emitting region should be larger than $(3^{4}L_{X}k^4/4{\pi}{\sigma}E^4_{\rm cut})^{1/2}$, which yields a value of $>50$\,km for the cutoff energy of the observed spectrum ($E_{\rm cut}{\sim}6$, see Sect.~3.1 and also \citealt{2018MNRAS.476L..45C}). 
The luminosity derived size and the  spectral fit derived size of $R_{\rm BB}{\sim}50$\,km, are unphysically high for an accretion column \citep{2015MNRAS.454.2539M}, thus challenging the basic principles of the AC model. On the contrary these sizes are in good agreement with the inner part of the magnetospheric envelope with the highest optical depth (see Fig. \ref{fig:envelope}).

Lastly, we note that the existence of the two areas (with and without the absorption feature) in the MCAE parameter space -- as revealed by the BXA runs (e.g. row 3 column 4 Fig.~\ref{fig:BXA_hist_MCD} -- signifies that the redundancy of the cyclotron-like feature may not be limited within just the MCAE model, but a set of models with a more complex continuum. Namely, it appears that the cyclotron feature is required by the model for a high k${\rm T_{hot}}$, resembling the smooth continuum of the AC model. The BXA results suggest that there are two classes of solutions to the problem of the ${\sim}10\,$keV residual structure. A smooth continuum that requires a broad absorption-like feature or a more complex continuum comprised of more than one component for the high-energy part.

In our analysis we explored the MCAE model. However, a more complex spectral continuum is also predicted for the broadband emission of the accretion column, as a result of bulk and thermal comptonization of thermal and Bremsstrahlung photons \citep[e.g.][]{2007ApJ...654..435B,2012MNRAS.424.2854F,2016A&A...591A..29F}. As these models consider the accretion column emission in X-ray pulsars below the ULX luminosity, they do not take into account the optically thick structure, expected to form around the accretion column and in the surrounding magnetosphere. For this reason we  considered the MCAE model as an alternative to a smooth cutoff power-law continuum. 
The MCAE model is a better description of the data according to the Bayesian model comparison.

 \section{Conclusions}
 We have carried out phase-resolved, broadband spectroscopy of \ulx. Our analysis reveals the presence of an accretion disk, indicated by the presence of non-pulsating soft thermal emission, consistent with a disk truncated at  approximately the magnetospheric radius of a highly magnetised NS. We further find that an additional, hard emission component is responsible for the observed pulsations. The flux of this component strongly varies with the pulse-phase but not significantly with its spectral hardness. 
 
 We find indications for the presence of a broad absorption-like feature that in the context of a highly magnetized neutron star may be interpreted as electron cyclotron line. If real, that would be the first such feature detected in an pulsating ULX, providing some constraint on the B-field of such a source. However, we find a strong dependence between the presence of the line and the choice of the underlying continuum model. To further probe this dependece we considered two different models for the source continuum, the AC and MCAE. Indeed, if the source continuum is modeled using the recently suggested MCAE model \citep{2017MNRAS.tmp..143M,2017A&A...608A..47K} for the spectra of (pulsating) ULXs, an absorption line is not required in order to successfully model the source spectrum. Therefore, the presence of the cyclotron line must be considered with great caution.
 
 To further probe the suitability of each of the two models we employed Bayesian techniques for model comparison and also qualitatively evaluated the physical characteristics of the source as inferred from the two models. The Bayesian analysis demonstrated that the MCAE model is preferable to the AC model and that -- while the presence of the cyclotron line in case of the MCAE model cannot be ruled out -- there is a well defined region in the MCAE parameter space, within which the cyclotron line is not detected.  Furthermore, the Bayesian model comparisons suggest that in the context of the MCAE, the model without the line is more likely, than the one that includes it. The cyclotron line is detected with much higher significance in the context of the AC model which is observed in most XRPs. However, if the observed emission originates in an accretion column questions are raised with respect to the very low energy of the roll-off, the shape of the pulse profile, the relation between the spectral slope and the pulse phase and the inferred size of the emitting region.
 
 All these source characteristics can be better understood in the context of the MCAE model, within which the presence of a cyclotron line is less probable. A third possibility involves the existence of the CRSF in the MCAE framework, in which case the inferred magnetic field strength on the surface of the NS is estimated to be higher than $10^{13}$\,G. However, the presence of such a value would be quite surprising in this context, due to the obscuring effects of the accretion envelope, but also in the AC model context if the origin of the cyclotron emission is due to reflection off the NS surface.

Confirmation of the presence of the cyclotron feature in \ulx is of great importance, as it will not only provide an indication of the magnetic field strength, but also provide crucial insight on the shape of the underlying continuum, which is still a matter of considerable debate for the ULX population. Therefore, deeper \nus observations of the source are strongly encouraged.

\begin{acknowledgements}
The authors would like to thank the anonymous referee whose contribution, significantly improved our manuscript. FK acknowledges support from the CNES. JB acknowledges support from the CONICYT-Chile grants Basal-CATA PFB-06/2007, FONDECYT Postdoctorados 3160439, and the Ministry of Economy, Development, and Tourism's Millennium Science Initiative through grant IC120009, awarded to The Millennium Institute of Astrophysics, MAS. 
\end{acknowledgements}

\bibliographystyle{aa} 
\bibliography{general} 

\begin{appendix}

\section{BXA analysis \ulx}
\label{BXA-spec}

\begin{table}
 \caption{Log-Evidence for the MCAE and AC models as described in Sect.~\ref{sec:Bayesian}.}
 \begin{center}
\scalebox{0.8}{   \begin{tabular}{lcc}
     \hline\hline\noalign{\smallskip}
      \multicolumn{3}{l}{MCAE model} \\
      \noalign{\smallskip}\hline\noalign{\smallskip}
        Phase & $\log{\rm Z}$ & DOF \\
     \noalign{\smallskip}\hline\noalign{\smallskip}
     Full (line)    & -461.5$\pm$0.3  & 761 \\
     Full (no line) & -460.62$\pm$0.28  & 764\\
      \noalign{\smallskip}\hline\noalign{\smallskip}         
      \noalign{\smallskip}\hline\noalign{\smallskip}
     \multicolumn{3}{l}{ cPL+dBB  (AC model)} \\
      \noalign{\smallskip}\hline\noalign{\smallskip}
        Phase & $\log{\rm Z}$ & DOF \\
     \noalign{\smallskip}\hline\noalign{\smallskip}
     Full (line)    & -467.0$\pm$0.3  &  762\\
     Full (no line) & -516.0$\pm$0.3 &  765\\
      \noalign{\smallskip}\hline\noalign{\smallskip}
\noalign{\smallskip}\hline\noalign{\smallskip}
     \multicolumn{3}{l}{ hec*PL+dBB } \\
      \noalign{\smallskip}\hline\noalign{\smallskip}
        Phase & $\log{\rm Z}$ & DOF \\
     \noalign{\smallskip}\hline\noalign{\smallskip}
     Full (line)    & -471.3$\pm$0.3  &  761\\
     Full (no line) & -493.7$\pm$0.3 &  764\\
      \noalign{\smallskip}\hline\noalign{\smallskip}      
      
    \end{tabular}   }
     \end{center}
    
 \label{tab:evidence2}
\end{table}

\begin{table*}
 \caption {Best fit values  for the broadband ${\chi}^2$ fit of combined \xmm and \nus observation of NGC300 ULX-1, for phase-average (FULL) spectra for the MCAE model {\it without} the Gaussian absorption line. The continuum was fitted separately with the \texttt{xspec} \texttt{diskbb} and \texttt{bbodyrad} (2nd row) models for the hot thermal emission.  All errors are in the 90$\%$ confidence range.}
 \begin{center}
\scalebox{0.9}{   \begin{tabular}{lcccccccccccc}
     \hline\hline\noalign{\smallskip}
         & & \multicolumn{2}{c}{dBB soft}  & \multicolumn{2}{c}{(d)BB hot}& \multicolumn{2}{c}{PL tail}& red.${\chi^2}$(dof)\\
      \noalign{\smallskip}\hline\noalign{\smallskip}
      \noalign{\smallskip}\hline\noalign{\smallskip}
     \multicolumn{1}{l}{} &
     \multicolumn{1}{c}{nH} &
     \multicolumn{1}{c}{k${\rm T_{disk}}$} &
     \multicolumn{1}{c}{${\rm {K_{disk}}}$} &
     \multicolumn{1}{c}{k${\rm T_{hot}}$} &
     \multicolumn{1}{c}{${\rm {{K_{hot}}}}$} &
     \multicolumn{1}{c}{${\rm \Gamma}$} &   
     \multicolumn{1}{c}{${\rm  {{K_{PL}}}}$} &  

          \multicolumn{1}{c}{}\\

     \noalign{\smallskip}\hline\noalign{\smallskip}
      
      \multicolumn{1}{c}{} &
      \multicolumn{1}{c}{[$10^{21}$\,cm$^{-2}$]} &     
      \multicolumn{1}{c}{keV} &
      \multicolumn{1}{c}{} &
      \multicolumn{1}{c}{keV} &
      \multicolumn{1}{c}{[${\times}10^{-3}$]} &
      \multicolumn{1}{c}{} &
      \multicolumn{1}{c}{[${\times}10^{-5}$]} &  
      \multicolumn{1}{c}{} \\
      \noalign{\smallskip}\hline\noalign{\smallskip}
           dBB    & 0.94$_{-0.09}^{+0.10}$& 0.29$\pm0.02$& 11.5$_{-2.43}^{+3.30}$ & 2.40$_{-0.05}^{+0.06}$ & 6.87$_{-0.48}^{+0.45}$ & 1.12$_{-0.49}^{+0.35}$ &2.31$_{-1.81}^{+4.32}$   & 1.10 (764)\\
      \noalign{\smallskip}\hline\noalign{\smallskip}
      \noalign{\smallskip}\hline\noalign{\smallskip}
         BB       & 1.13$_{-0.14}^{+0.13}$& 0.26$\pm0.03$& 10.7$_{-4.43}^{+8.50}$ & 1.56${\pm}{0.03}$      & 69.5$_{-3.41}^{+3.98}$ & 1.99${\pm}{0.05}$      &66.1$_{-6.59}^{+5.56}$   & 1.11 (764) \\
      \noalign{\smallskip}\hline\noalign{\smallskip}
      \noalign{\smallskip}\hline\noalign{\smallskip}
    \end{tabular}   }
 \end{center}
  \tablefoot{{ ${\rm K_{disk}}$ is the normalization parameter for the \texttt{ diskbb} component. Namely ${\rm K_{disk}}$ = ${\rm(R_{\rm disk}/{D_{10kpc}})^{2}}\,\cos{i}$ and ${\rm K_{hot}}$ = ${\rm(R_{\rm disk}/{D_{10kpc}})^{2}}$ for the \texttt{bbodyrad} model (2nd row), where $R_{\rm disk}$ is the inner radius of the disk in km, ${\rm D_{10kpc}}$ is the distance in units of 10\,kpc and i is the inclination. ${\rm K_{hot}}$ is the same as ${\rm K_{disk}}$ but for the hot MCD component. ${\rm {{K_{cPL}}}}$ and ${\rm {{K_{PL}}}}$ are the power-law normalization parameters in units of photons/keV/cm$^2$/s at 1\,keV.  \\}  }
 \label{app2noline}
\end{table*}

\begin{figure*}
    \resizebox{\hsize}{!}{
     \includegraphics[angle=0,clip=]{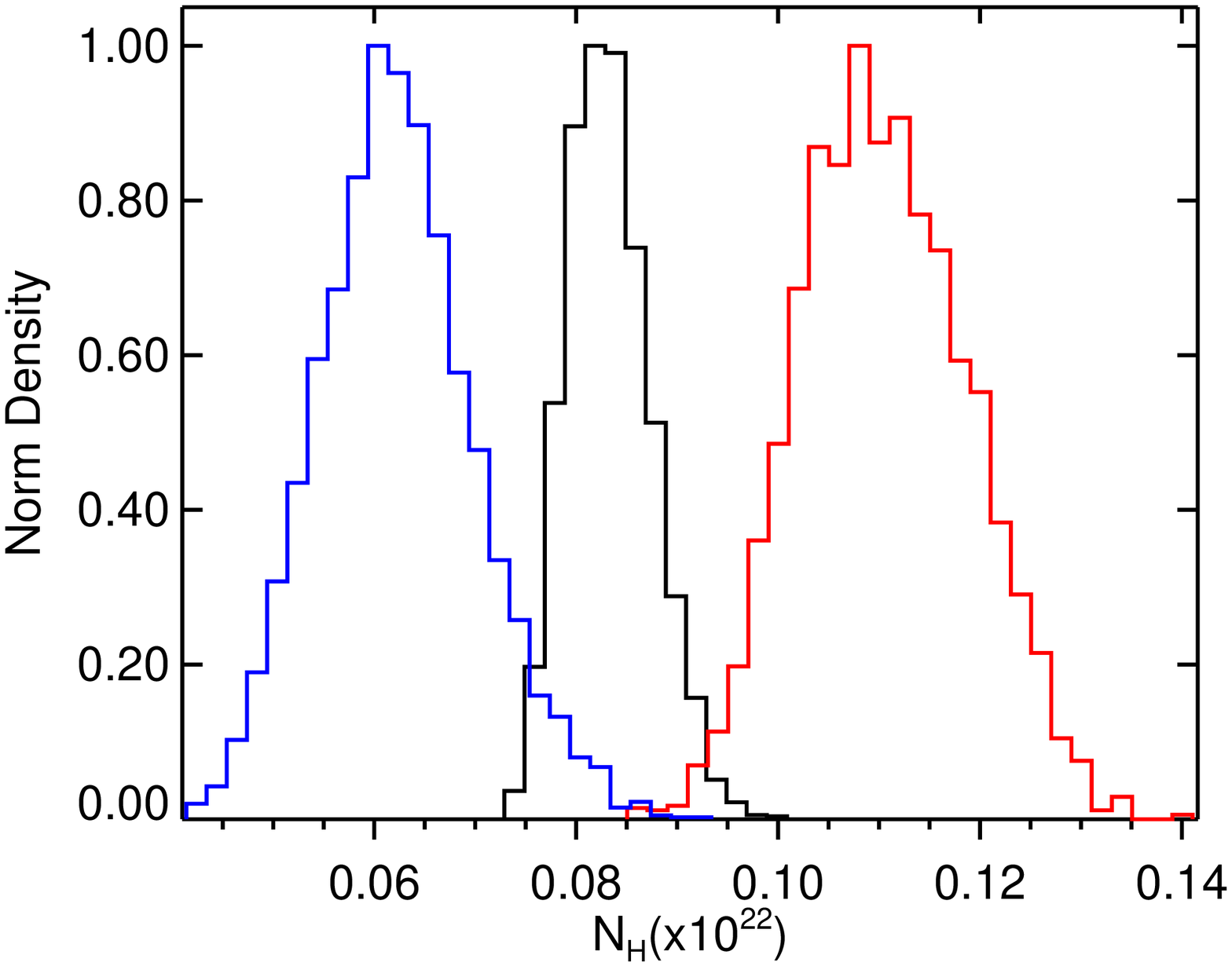}
     \includegraphics[angle=0,clip=]{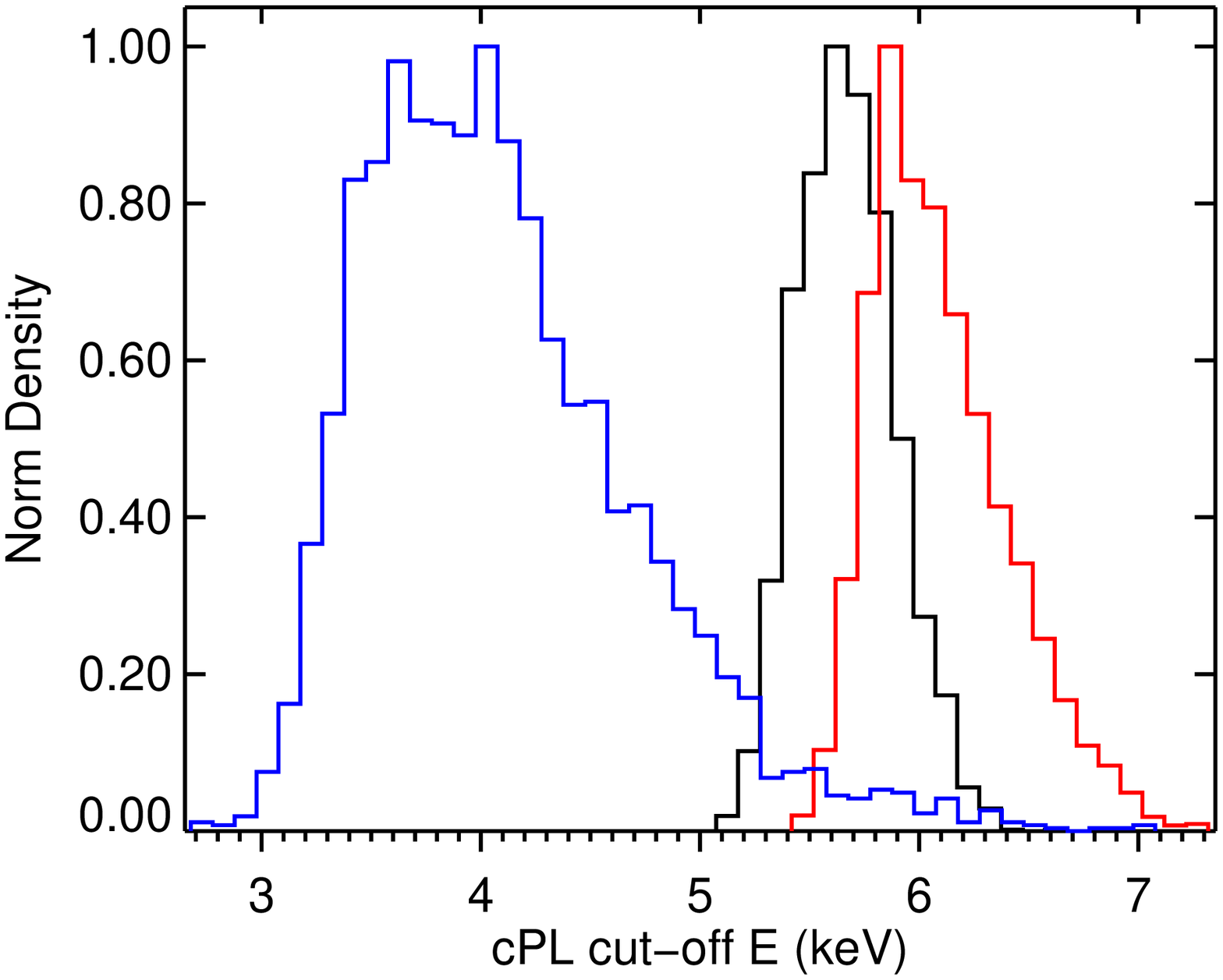}
     \includegraphics[angle=0,clip=]{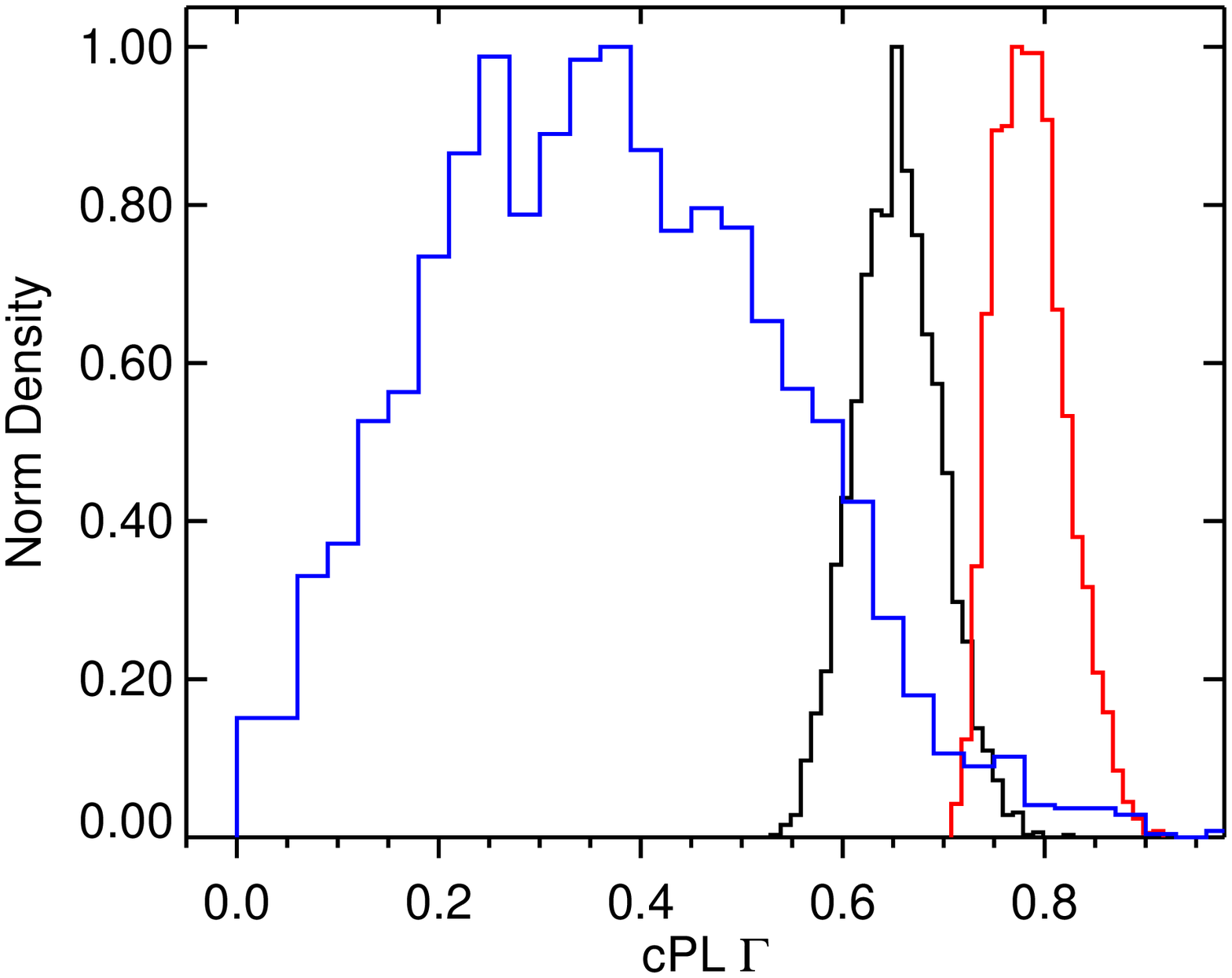}
     }
    \resizebox{\hsize}{!}{
     \includegraphics[angle=0,clip=]{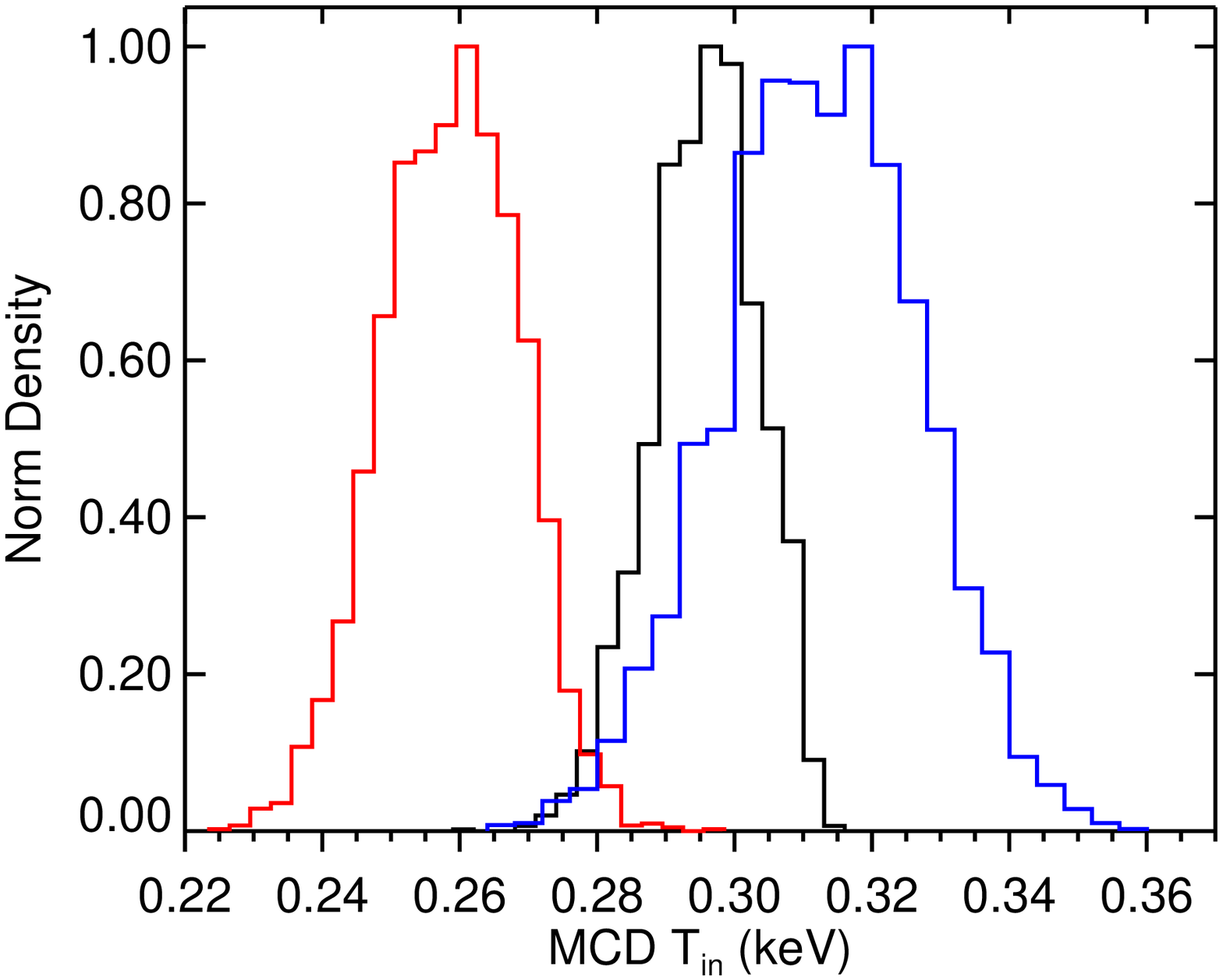}
     \includegraphics[angle=0,clip=]{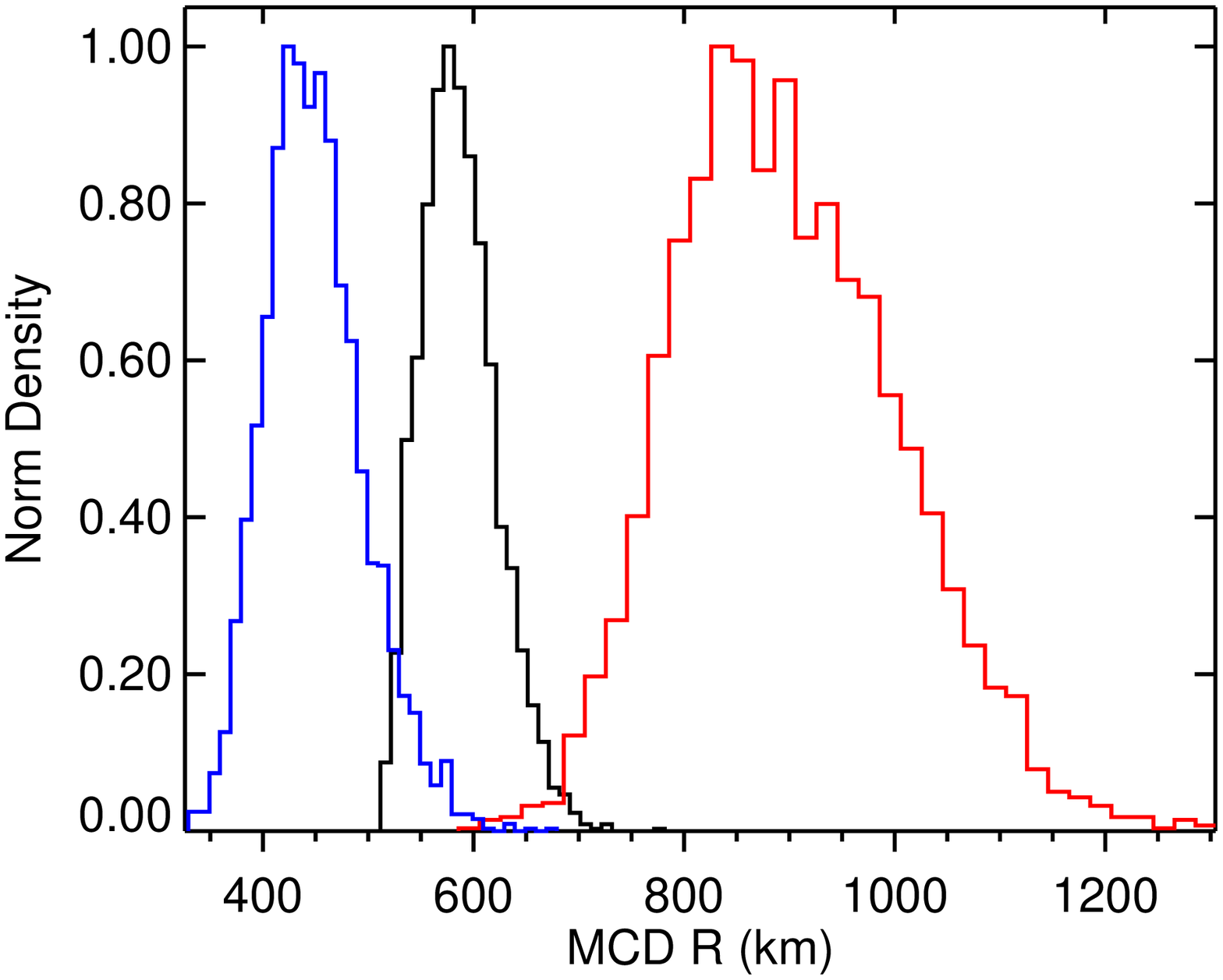}
     \includegraphics[angle=0,clip=]{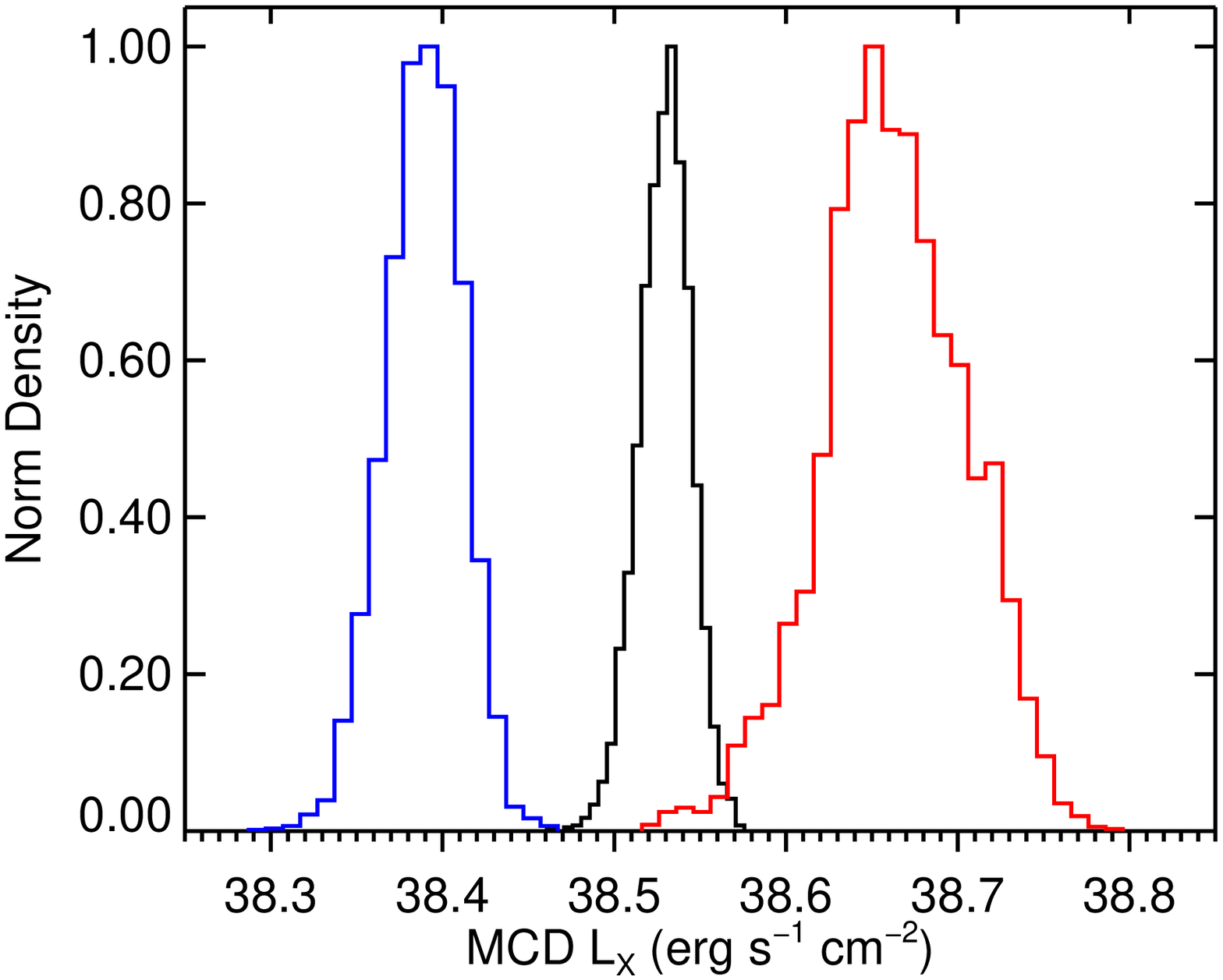}
     }
    \resizebox{\hsize}{!}{
     \includegraphics[angle=0,clip=]{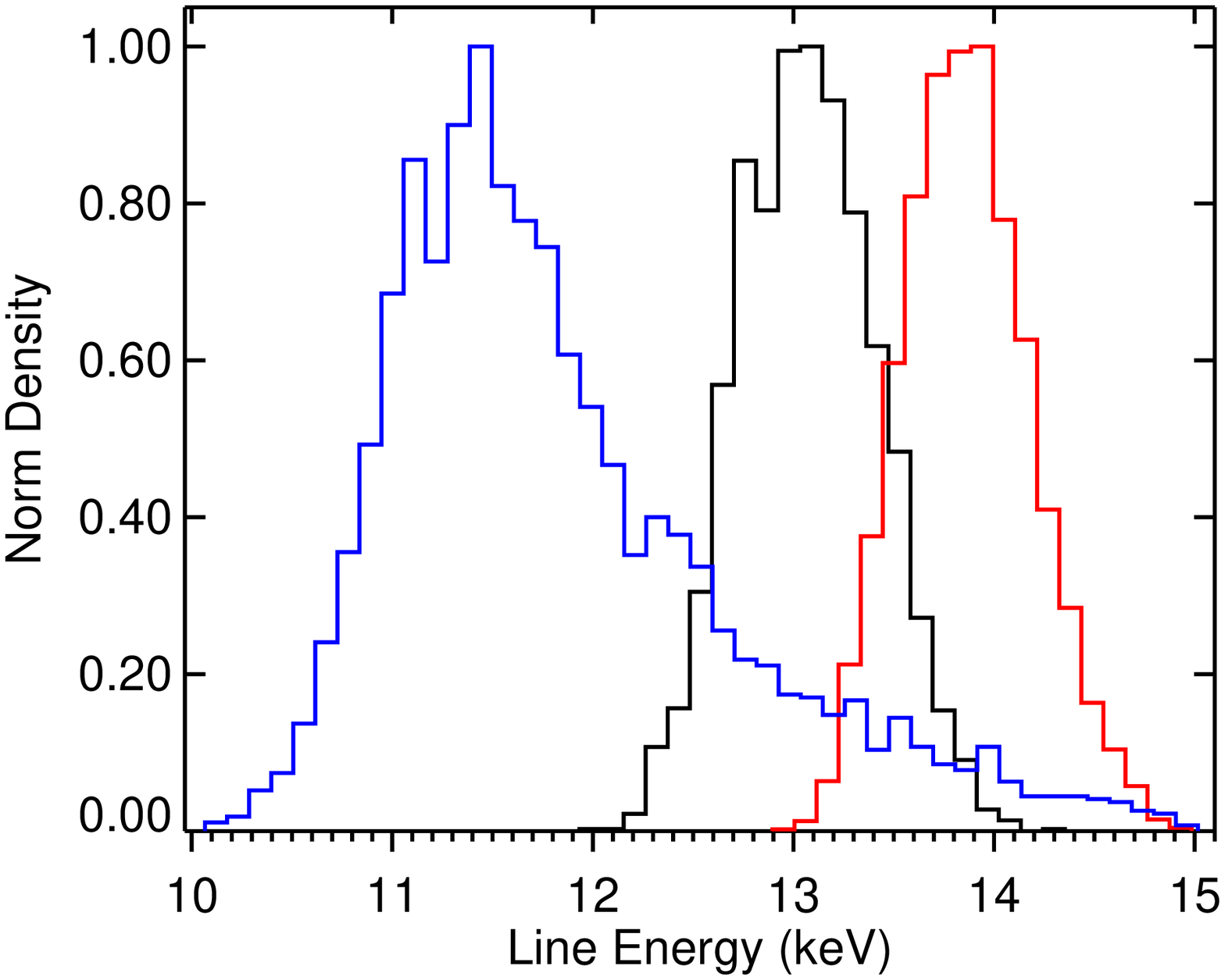}
     \includegraphics[angle=0,clip=]{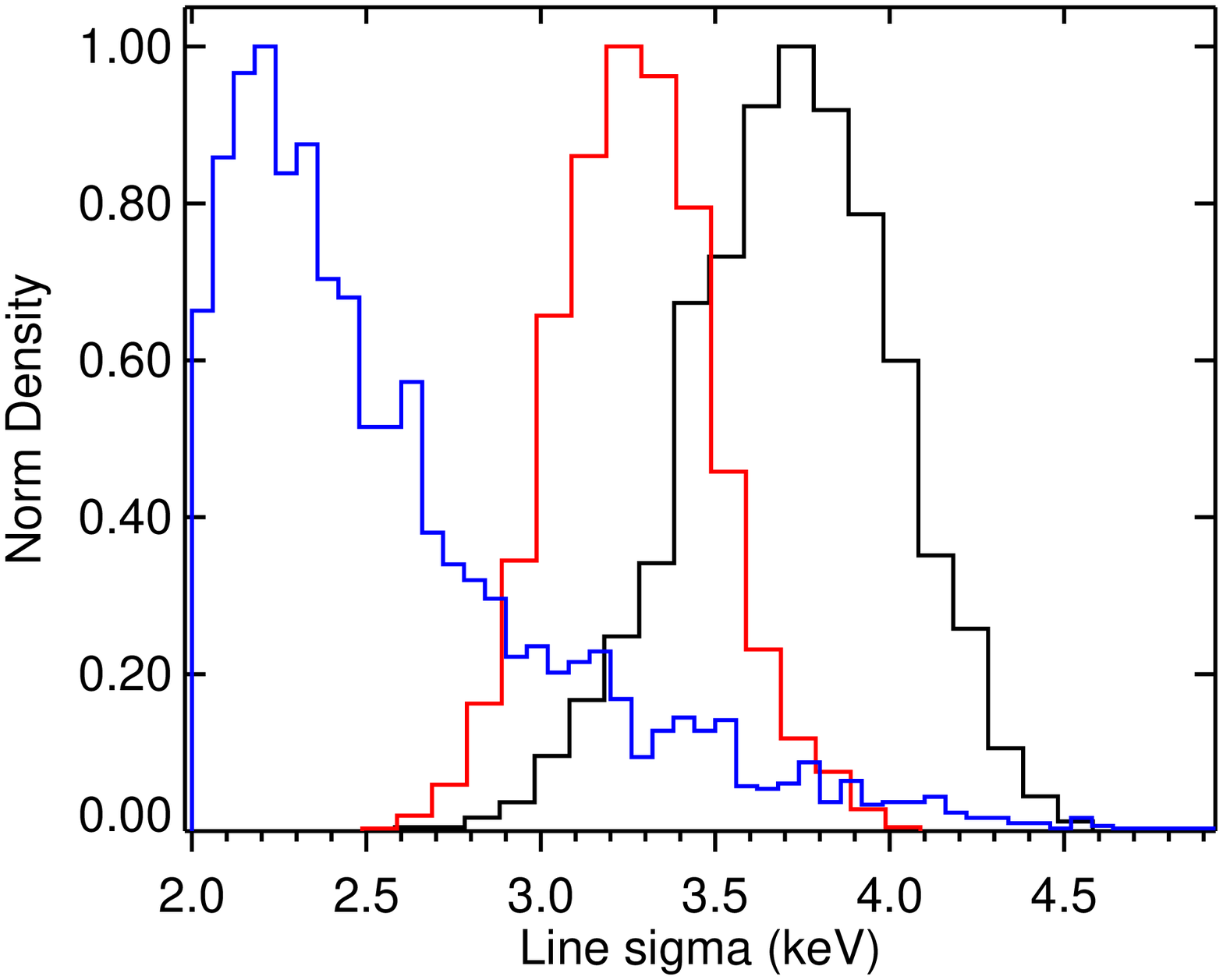}
     \includegraphics[angle=0,clip=]{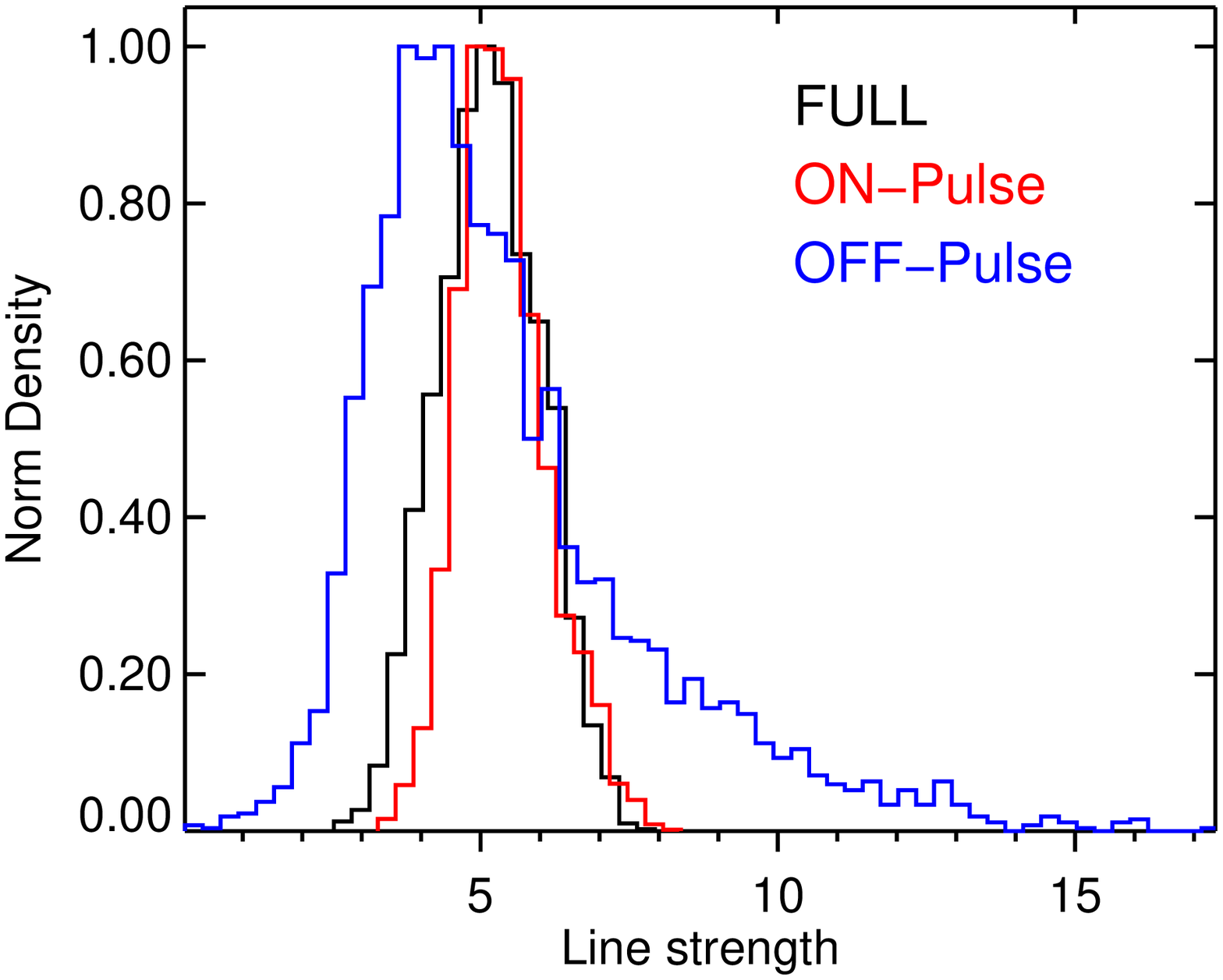}
     }
  \caption{Marginalised parameters of the AC model. The pannels contain the posterior probability density distributions, normalised to the maximum, for the phase averaged (black lines), ON-Pulse (red lines) and OFF-Pulse (blue lines) spectra.  The red lines mark the 5\%, 50\% (cross-section) and 95\% percentiles of the 1D distributions.  }
   
  \label{fig:BXA_hist}
\end{figure*}

\begin{figure*}
\resizebox{\hsize}{!}{
     \includegraphics[angle=0,clip=]{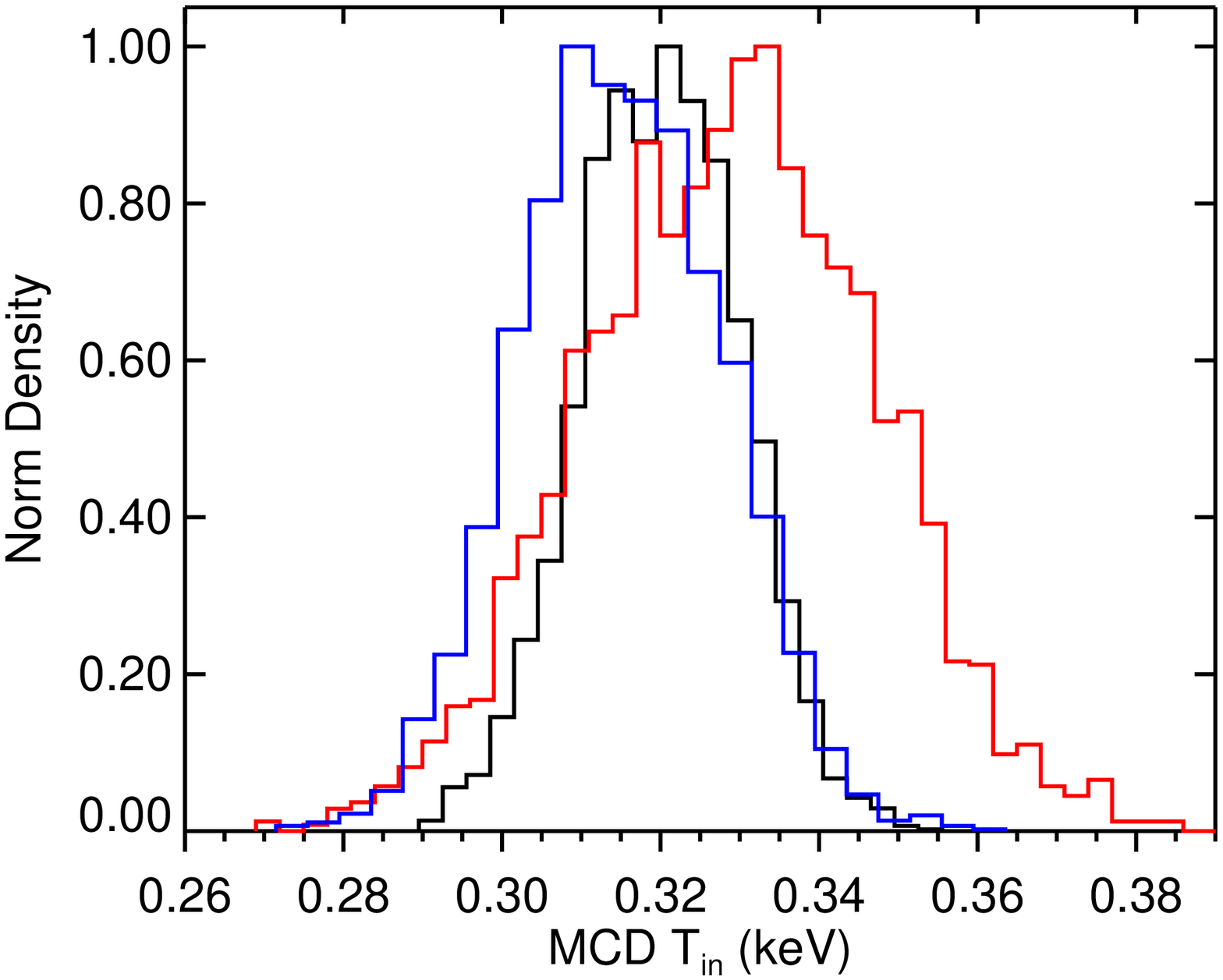}
     \includegraphics[angle=0,clip=]{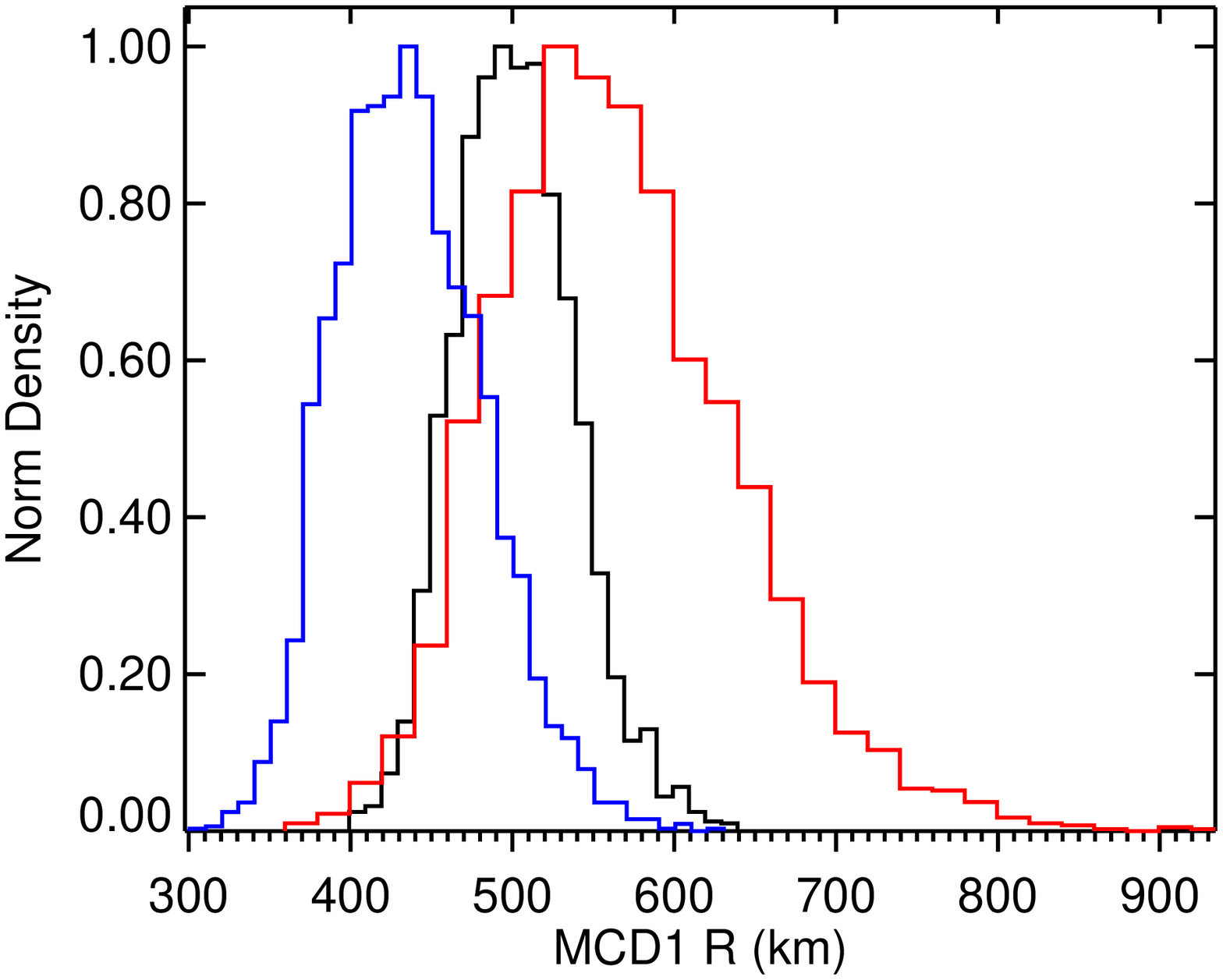}
     \includegraphics[angle=0,clip=]{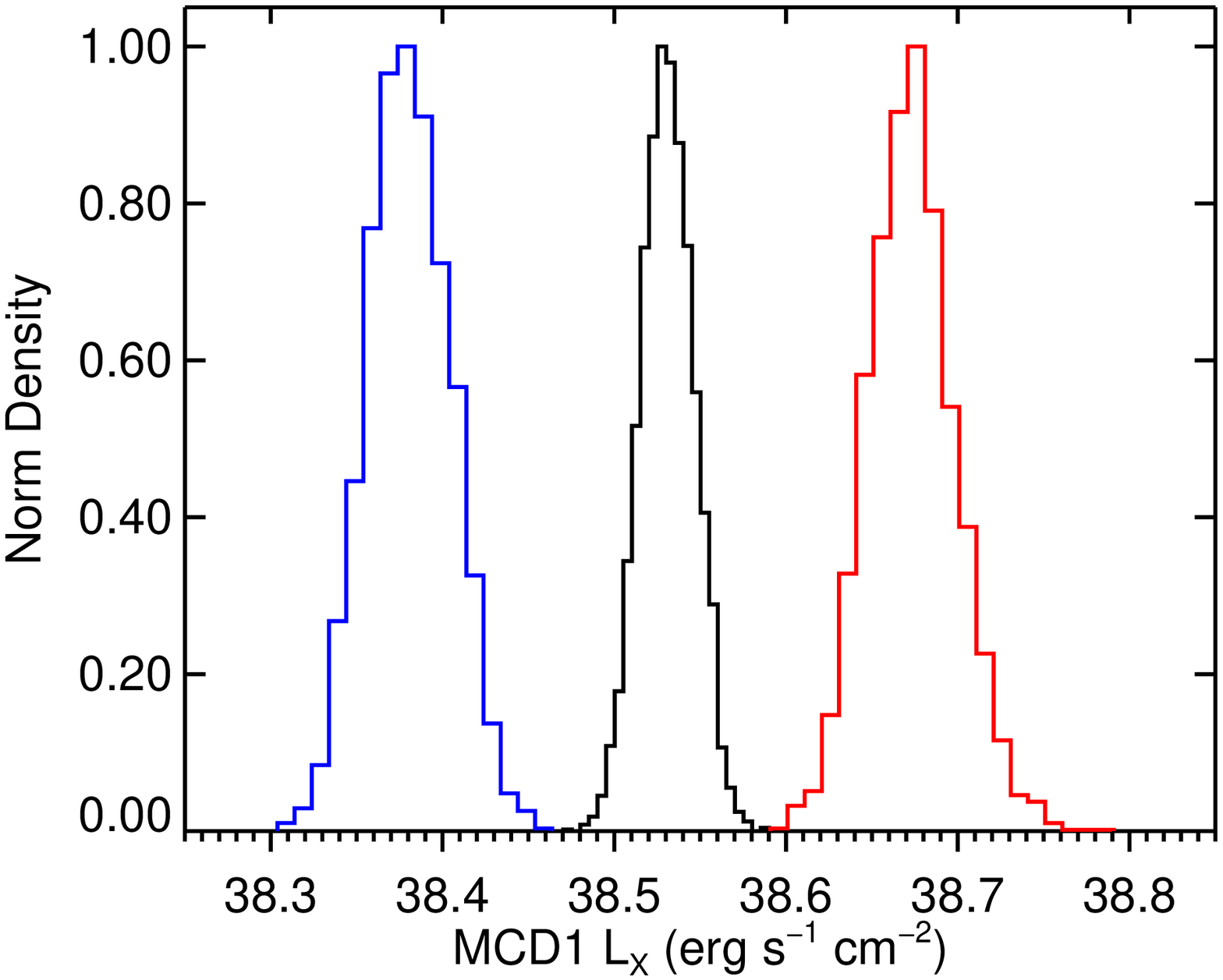}
     }
\resizebox{\hsize}{!}{
     \includegraphics[angle=0,clip=]{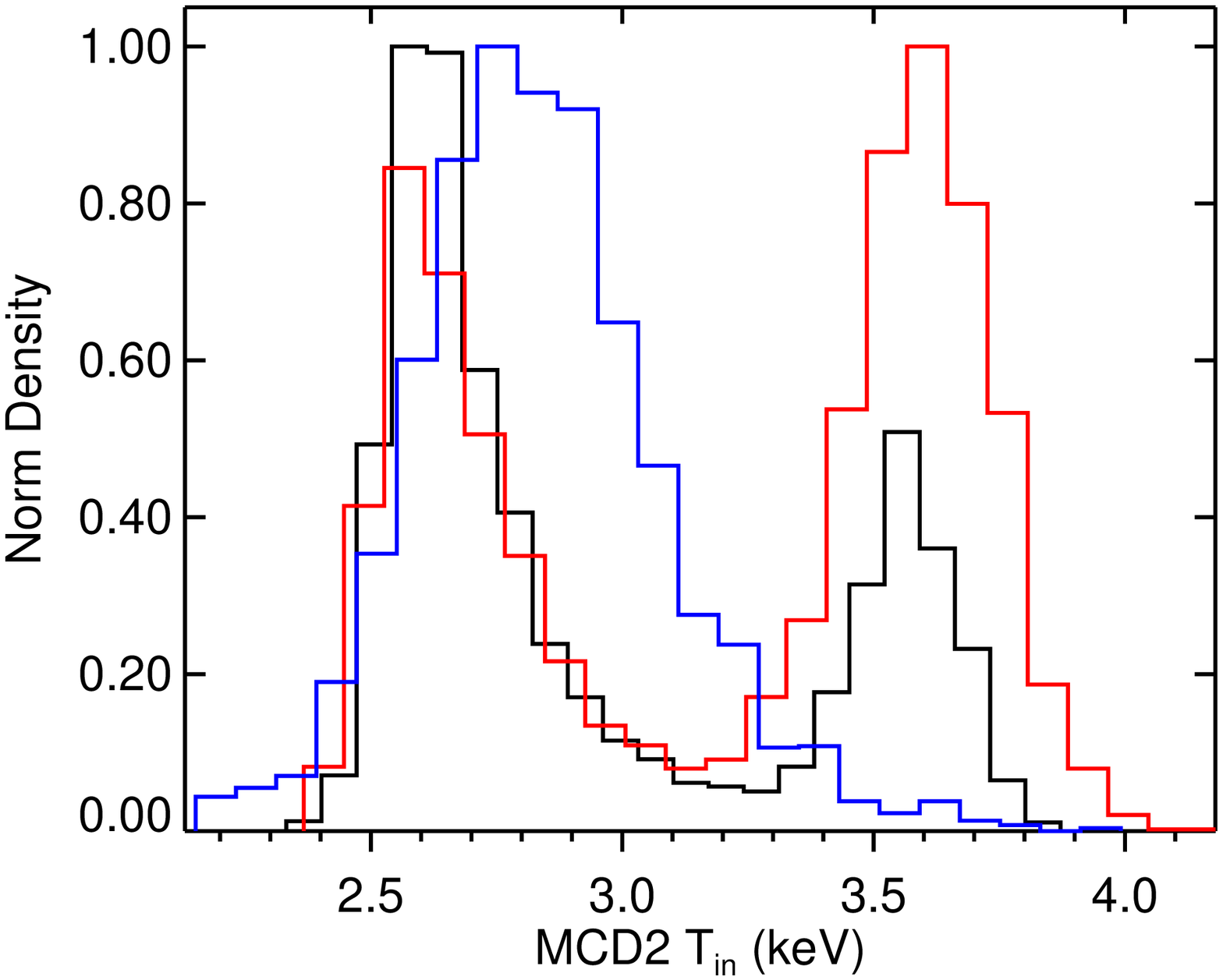}
     \includegraphics[angle=0,clip=]{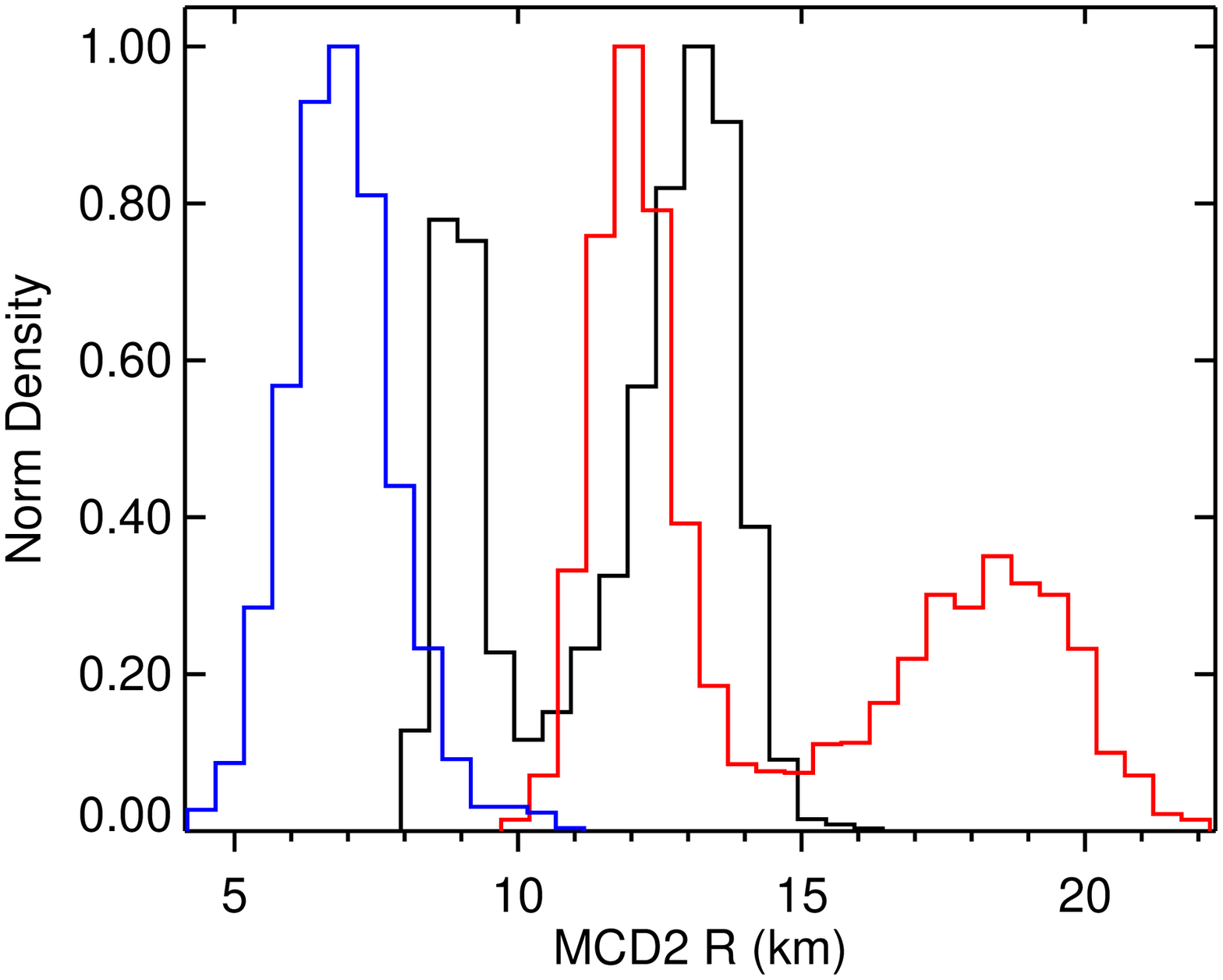}
     \includegraphics[angle=0,clip=]{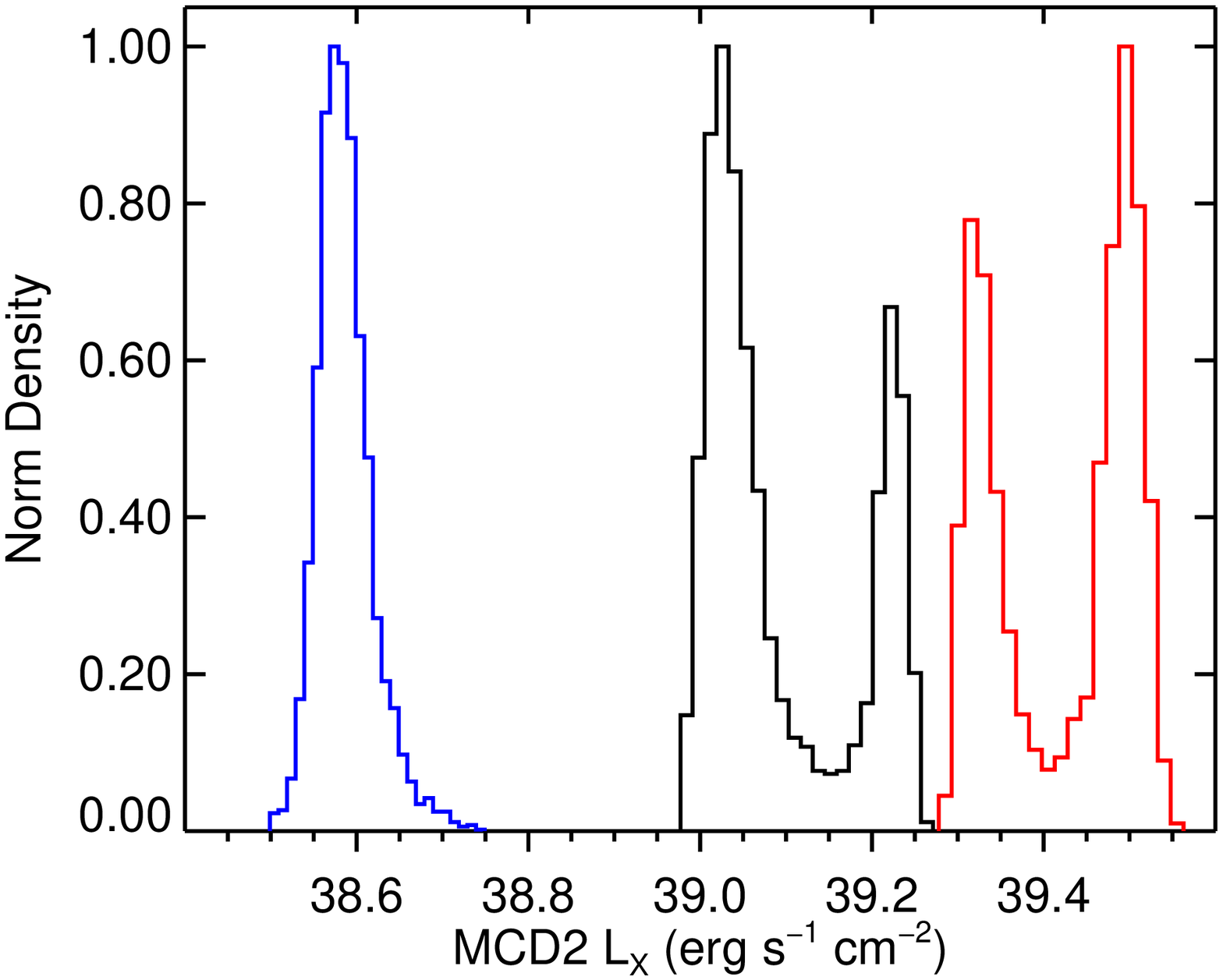}
     }
\resizebox{\hsize}{!}{
     \includegraphics[angle=0,clip=]{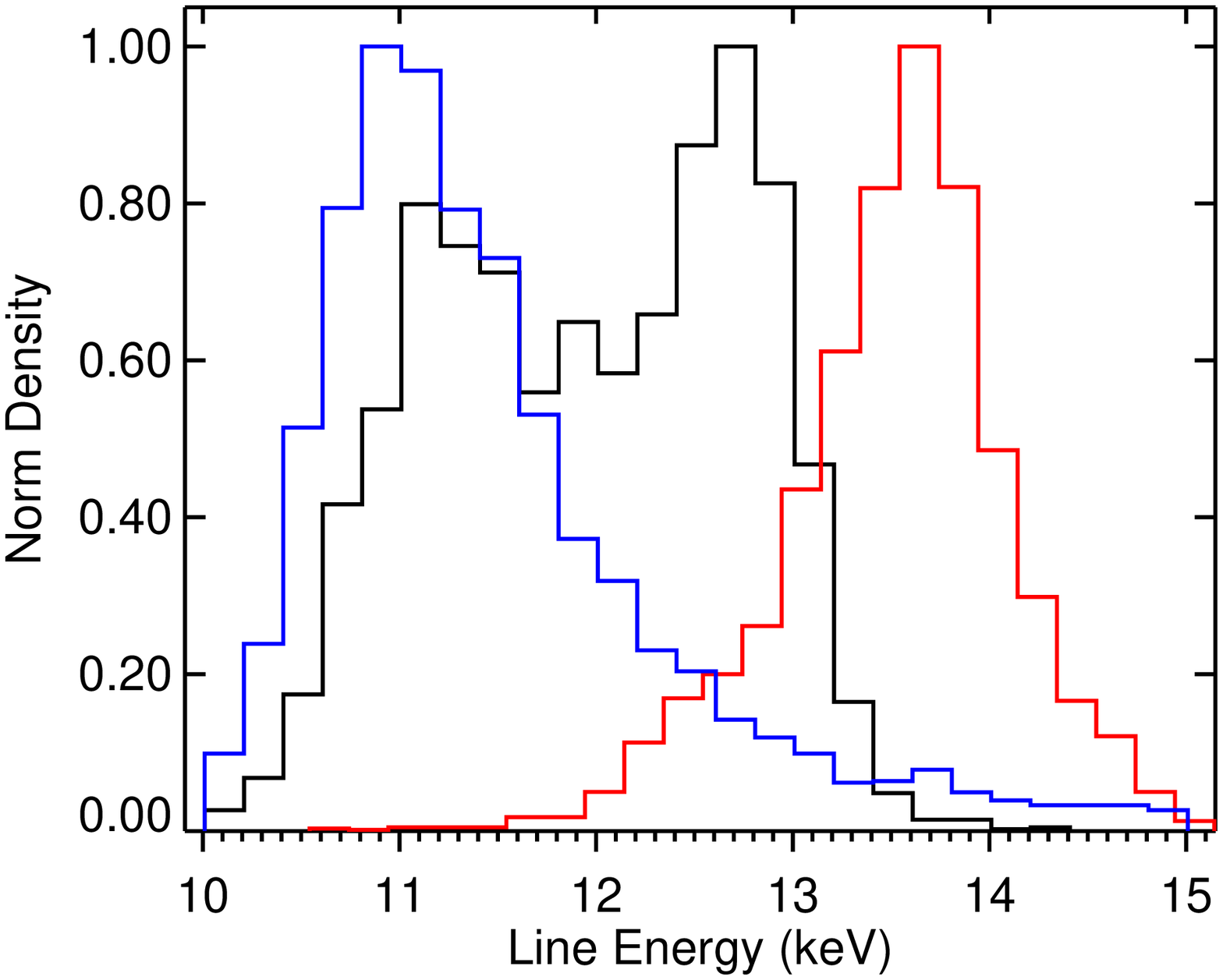}
     \includegraphics[angle=0,clip=]{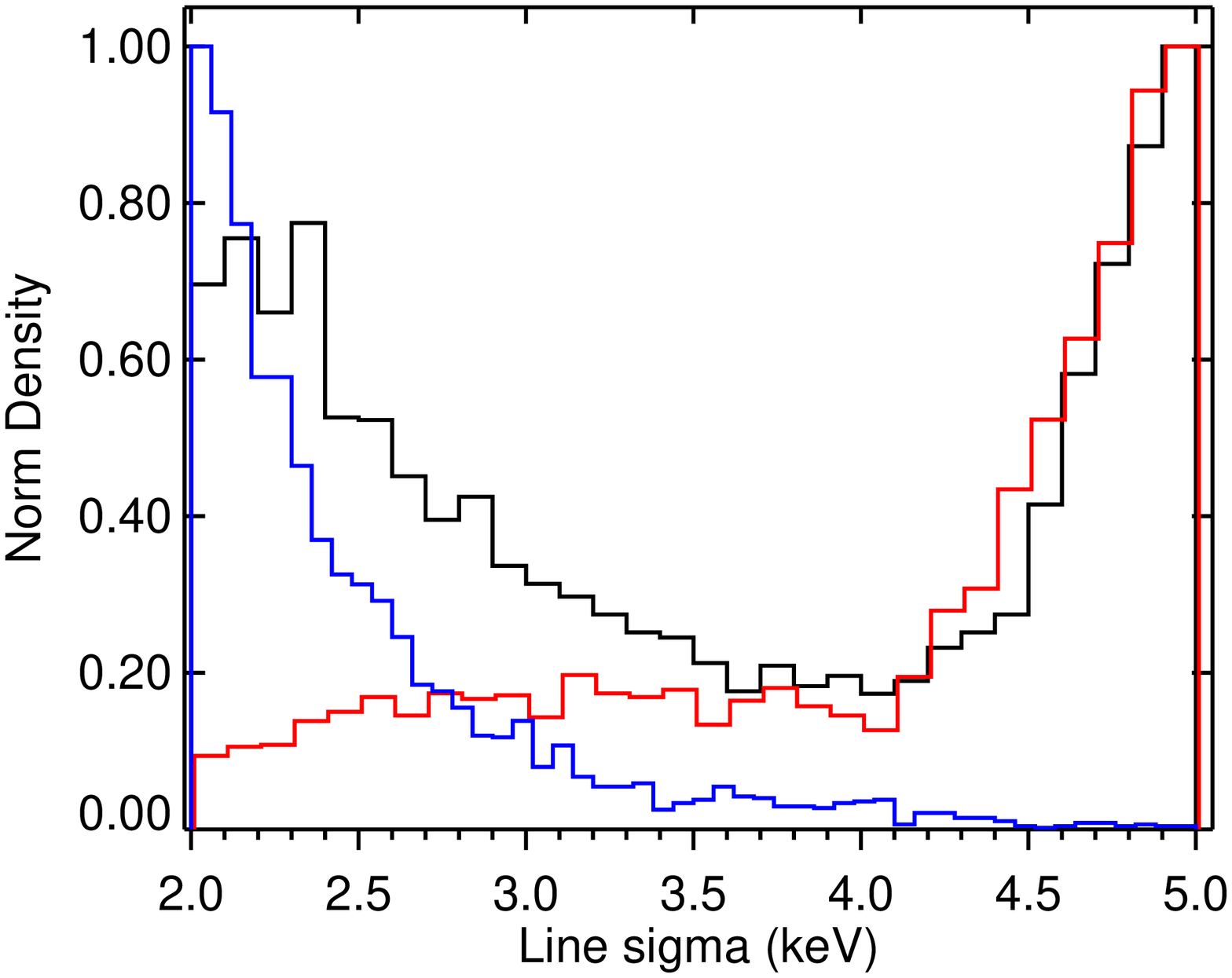}
     \includegraphics[angle=0,clip=]{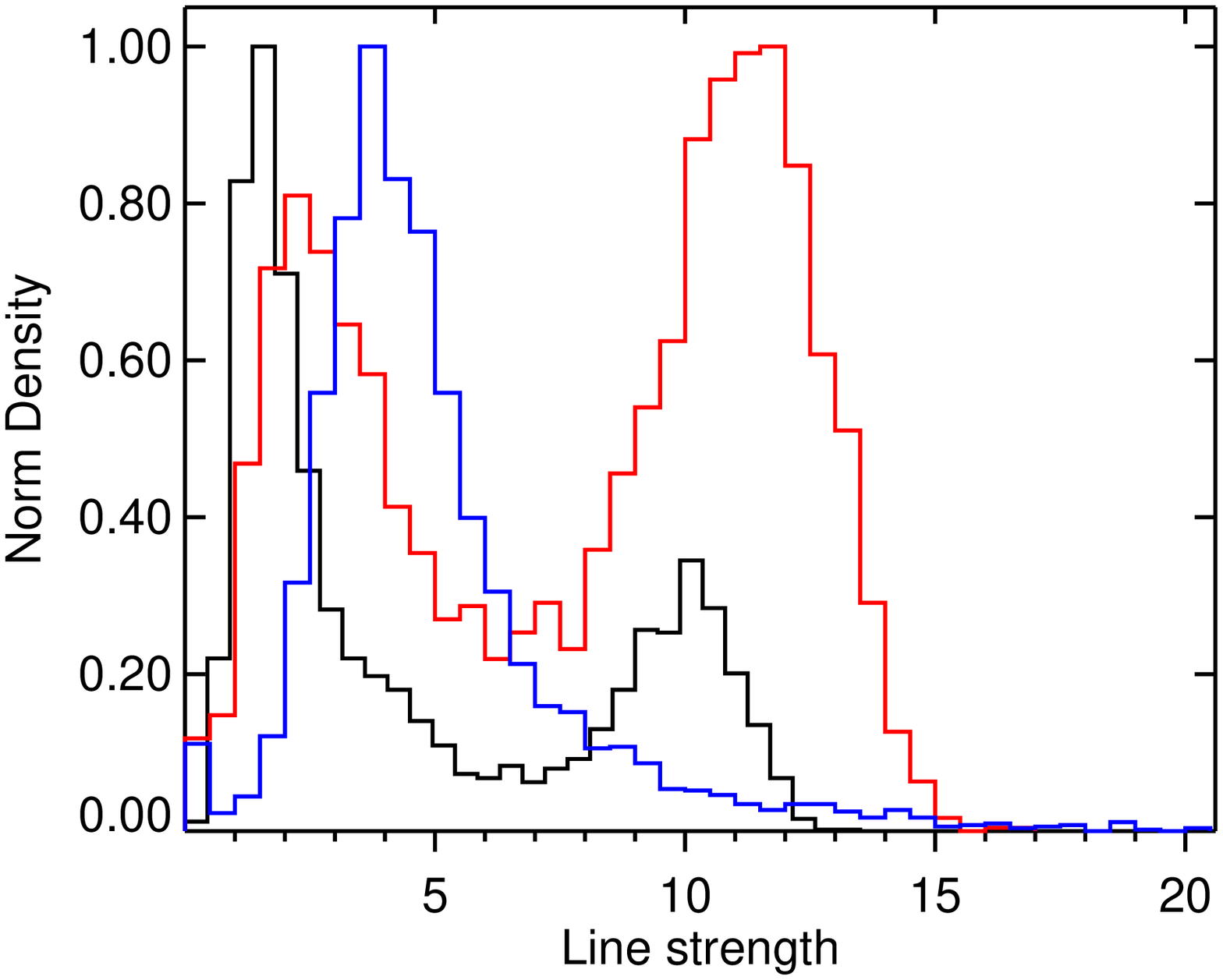}
     }
\resizebox{\hsize}{!}{
     \includegraphics[angle=0,clip=]{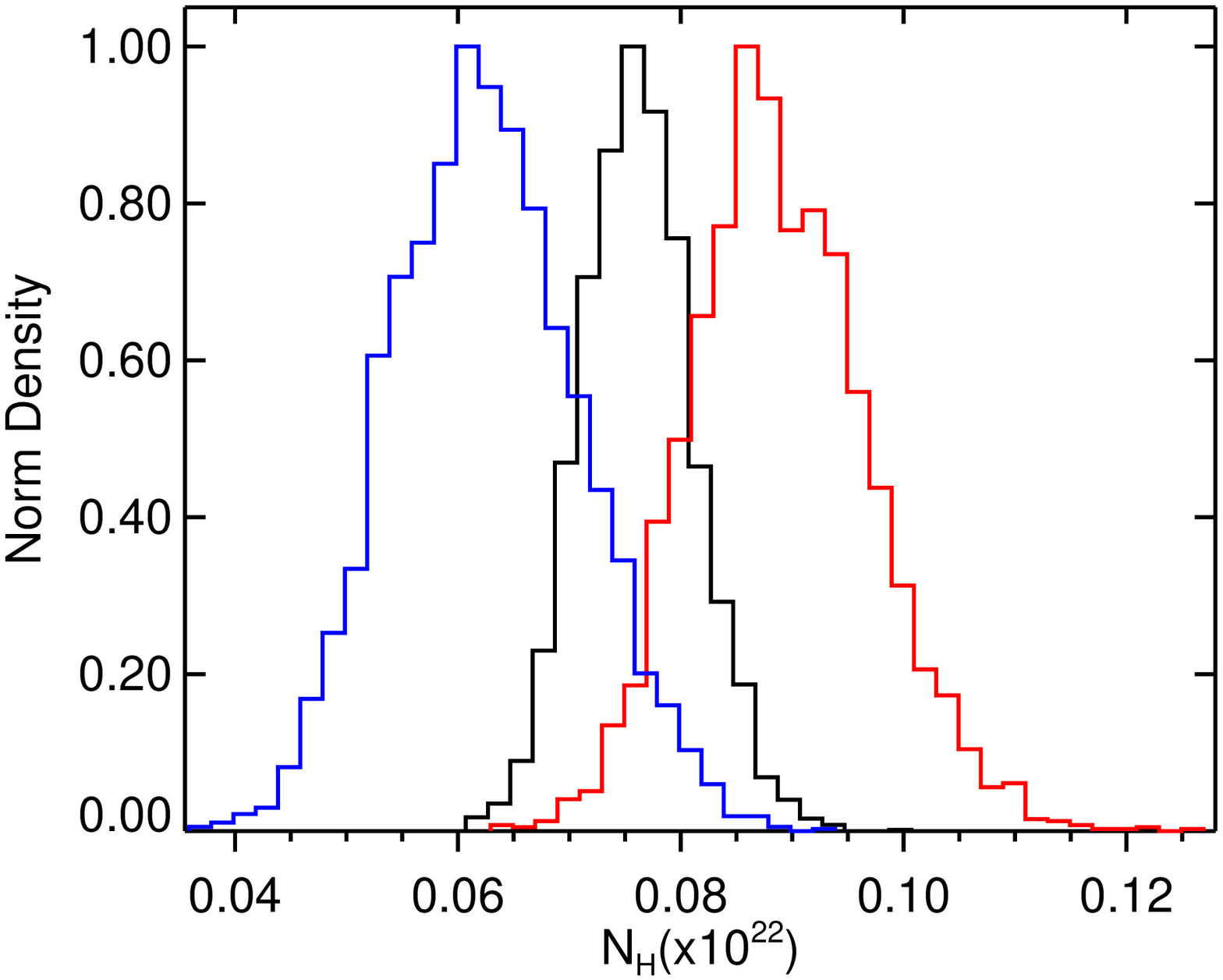}
     \includegraphics[angle=0,clip=]{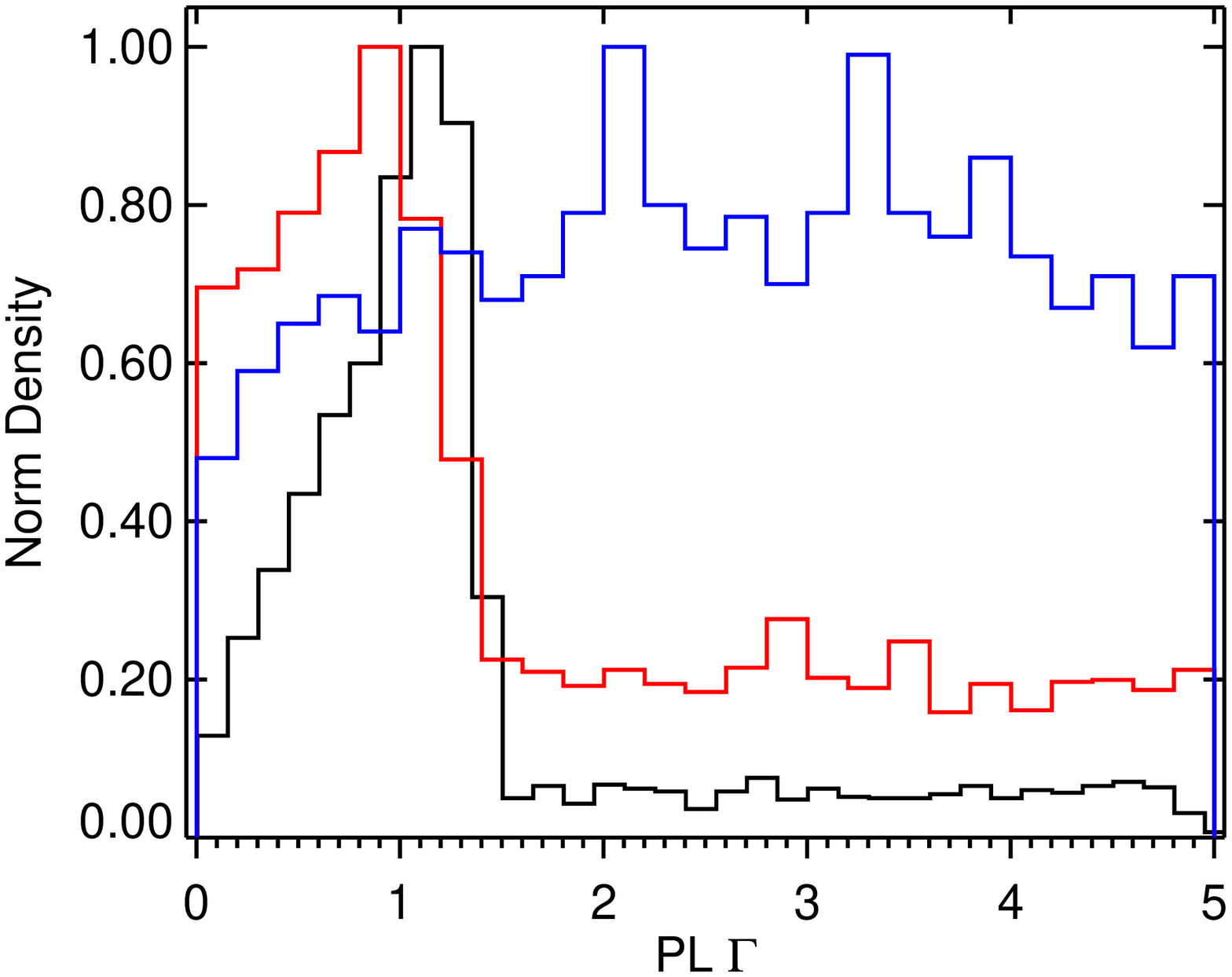}
     \includegraphics[angle=0,clip=]{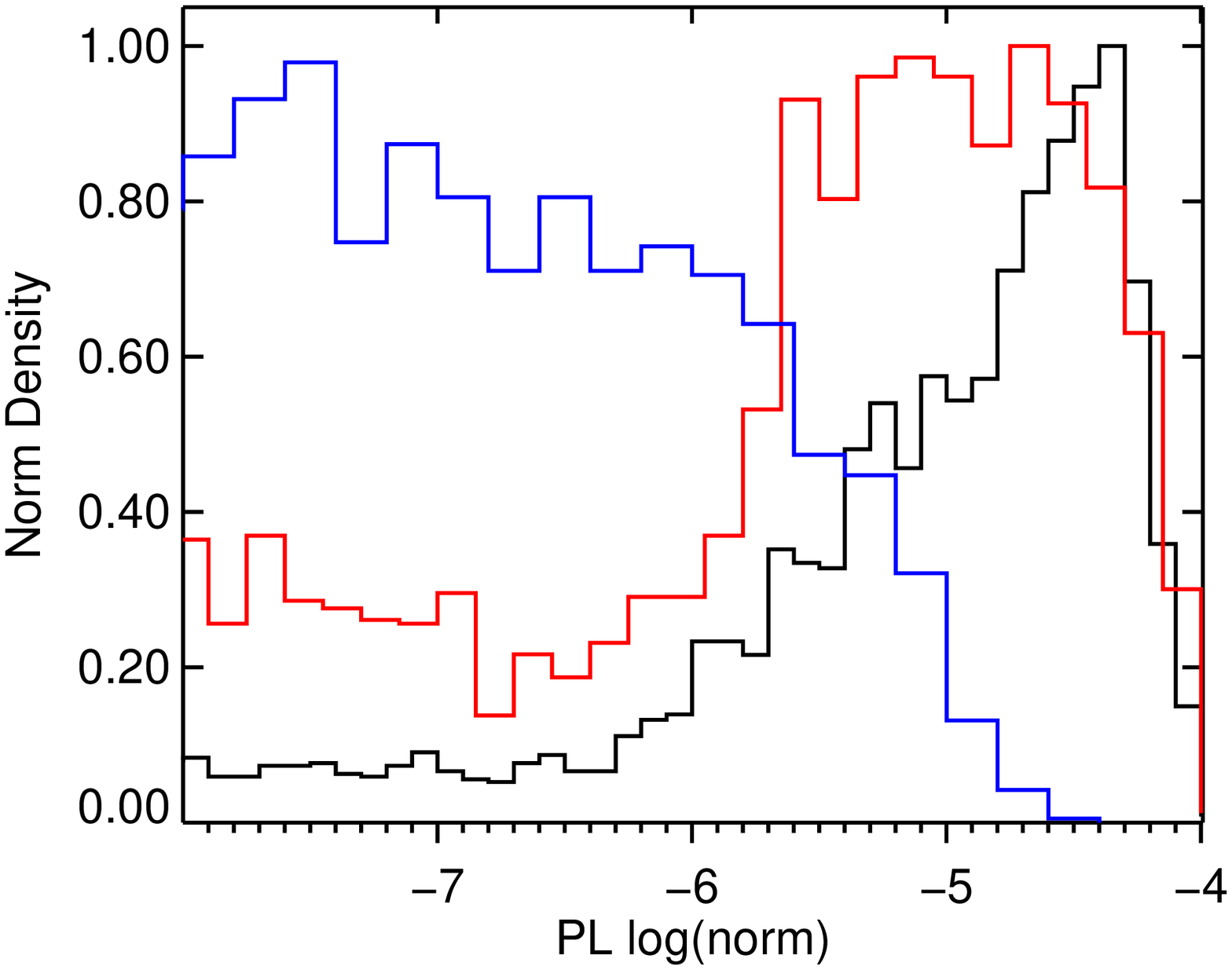}
     }
  \caption{Same as Fig. \ref{fig:BXA_hist}, but for MCAE model.} 
  \label{fig:BXA_hist_MCD}
\end{figure*}

\begin{figure*}
  \resizebox{\hsize}{!}{
     \includegraphics[angle=0,clip=]{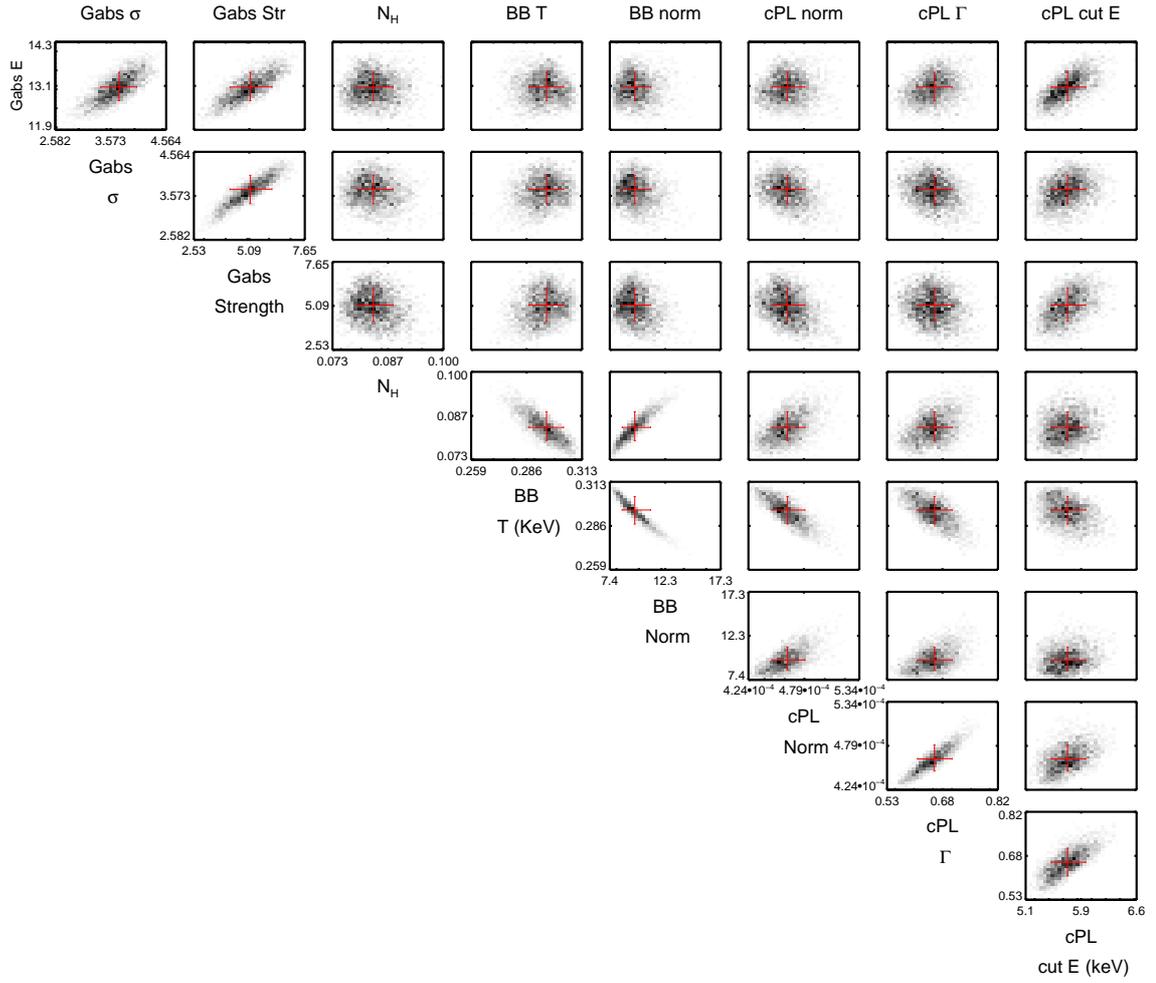}
     }
  \caption{Marginalised parameters of the AC model. For the distributions of Fig. \ref{fig:BXA_hist} we have plotted the 2D histograms for each pair of parameters. } 
  \label{fig:BXA_cpl}
\end{figure*}

\begin{figure*}
  \resizebox{\hsize}{!}{
     \includegraphics[angle=0,clip=]{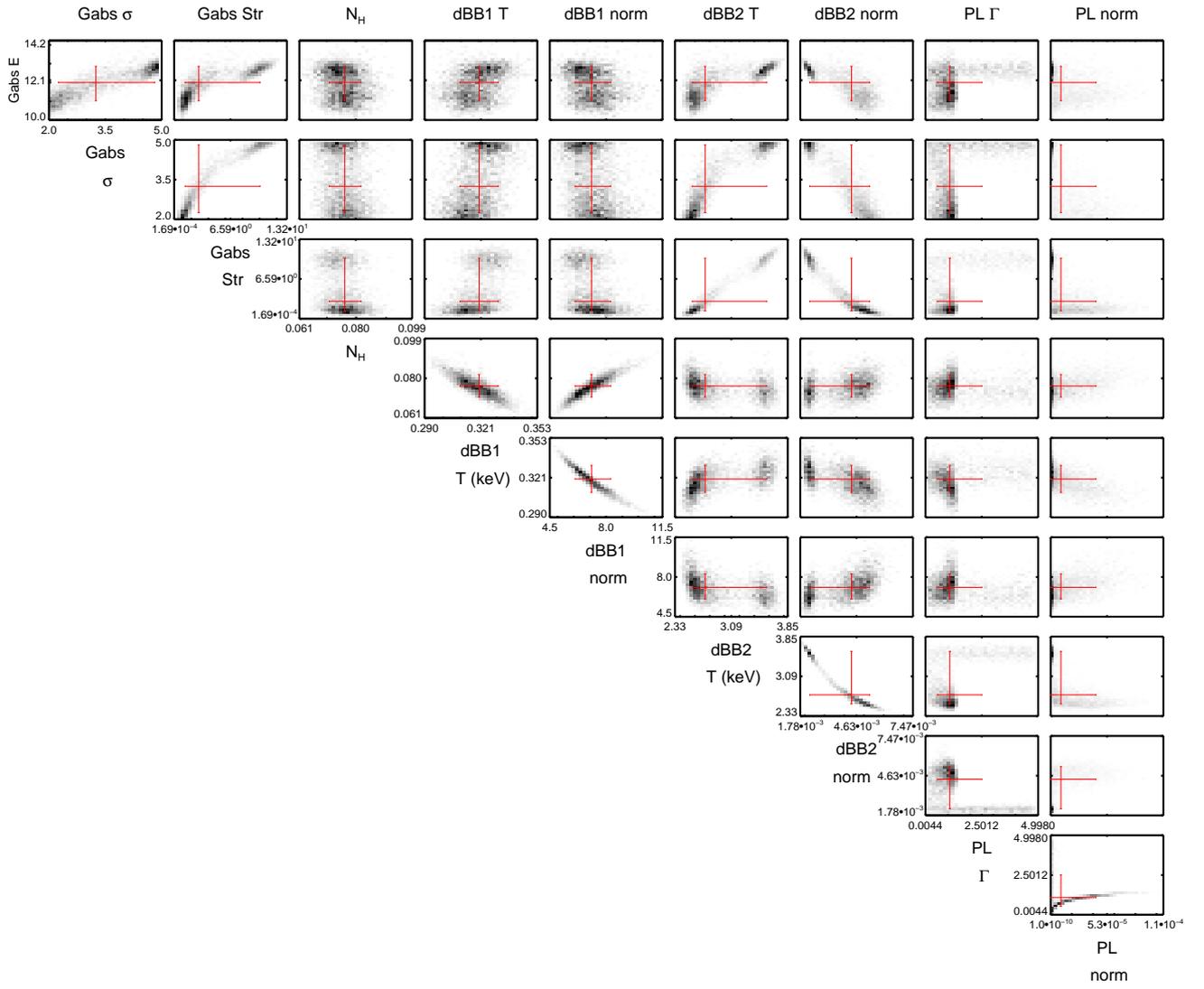}
     }
  \caption{Same as Fig. \ref{fig:BXA_cpl}, but for MCAE model.} 
  \label{fig:BXA_mcd}
\end{figure*}

\end{appendix}

\end{document}